\documentclass[twocolumn,letterpaper,pra,showpacs,superscriptaddress]{revtex4}
\usepackage{graphicx}% eps files
\usepackage{bm}% bold math
\usepackage{amsfonts}% equation formatting
\usepackage[latin1]{inputenc} 
\usepackage[intlimits]{amsmath}
\usepackage{amsxtra}
\usepackage{amssymb}
\usepackage{psfrag}
\usepackage{amstext}
\newlength{\twocolumnwidth}

\newlength{\auxlv}

\newlength{\DL}

\newlength{\pwd} 
\begin{document} 
%******************************************* 
\title{Causal signal transmission by quantum fields.%
\\ Phase-space approach to quantum electrodynamics.%
} 
%*******************************************
\author{L.\ I.\ Plimak} 
\affiliation{Institut f\"ur Quantenphysik, Universit\"at Ulm, 
D-89069 Ulm, Germany.} 
%*******************************************
\author{S.\ Stenholm} 
\affiliation{Institut f\"ur Quantenphysik, Universit\"at Ulm, 
D-89069 Ulm, Germany.} 
\affiliation{Physics Department, Royal Institute of Technology, KTH, Stockholm, Sweden.} 
\affiliation{Laboratory of Computational Engineering, HUT, Espoo, Finland.} 
%*******************************************
\date{\today} 
%*******************************************
\begin{abstract} 
%*******************************************
 
Phase-space techniques are generalised to nonlinear quantum electrodynamics beyond the rotating wave approximation, resulting in an essentially classical picture of radiation dynamics. 
%******************************************* 
\end{abstract}
%******************************************* 
\pacs{XXZ}
%*******************************************
\maketitle 
%*******************************************
%\tableofcontents 
%********************************************************
% hideV
%********************************************************

{\bf This is an early version of \cite{ClBehD}, with a number of formal details omitted there.\/}

%********************************************************
\section{Introduction}\label{ch:IntVI}%
%********************************************************
The goal of this paper is to generalise phase-space techniques to nonlinear quantum electrodynamics. A systematic introduction to conventional phase-space concepts may be found, e.g., in Mandel and Wolf \cite{MandelWolf}. For our methods we owe a lot to Agarwal and Wolf \cite{AgarwalWolf}. It is advisable for the reader to familiarise himself with 
section II 
of our paper \cite{QDynResp} before continuing with this one. 

Coherent states of the harmonic oscillator, which traditionally serve as an entry point to phase-space approaches, were introduced by Schr\"odinger as early as 1926 \cite{SchrCohSt}. That quantum dynamics of free bosonic systems maps to classical dynamics irrespective of the quantum state was firstly noticed by Feynman in his review on path integrals \cite{Feynman}. This understanding was instrumental in developing quantum theory of photodetection by Glauber \cite{GlauberPhDet}. Glauber's theory was initially formulated for free electromagnetic fields. It was extended to interacting fields by Kelley and Kleiner \cite{KelleyKleiner,GlauberTN}. However, de Haan \cite{deHaan} and later Bykov and Tatarskii \cite{BykTat,Tat} pointed out that Kelley-Kleiner's results are limited to the resonance, or rotating wave, approximation (RWA). Taking them outside the RWA leads to causality violations. In this paper, we lift this last restriction, generalising phase-space concepts to an arbitrary case of electromagnetic interaction. 

Our approach \cite{API,APII,APIII,WickCaus,QDynResp} (see also \mbox{Refs.\ \protect\cite{Belinicher,Corresp,BWO}}) hinges on explicit causality as a guiding principle. It originates in the observation that the commutator of free electromagnetic-field operators depends only on the linear response function, or, which is the same, retarded Green function, of the free field. In \mbox{Ref.\ \protect\cite{Corresp}} the formula relating commutator to response was called {\em the wave quantisation relation\/}. 
In \mbox{Ref.\ \protect\cite{API}}, we demonstrated that the wave quantisation relation leads to the so called {\em response transformation\/} of linear quantum dynamics. It was shown that the latter serves as an alternative entry point to conventional phase-space approaches, revealing the profound connection between causality (inherent in response of the field) and such formal concepts as normal ordering of bosonic operators. In \mbox{Refs.\ \protect\cite{APII,APIII}}, we postulated response transformations for Heisenberg field operators and showed that it results in a natural {\em response formulation\/} of quantum fields, which is at the same time a phase-space formulation generalised beyond the RWA. 

Analyses in \mbox{Ref.\ \protect\cite{API}} were limited to linear systems, while those in \mbox{Refs.\ \protect\cite{APII,APIII}} were to a large extent kinematical. Response transformation of nonlinear quantum dynamics was developed in \mbox{Refs.\ \protect\cite{WickCaus,QDynResp}}. In \mbox{Ref.\ \protect\cite{WickCaus}} we applied response transformation to Wick's theorem. The emerging relations were called {\em causal Wick theorems\/}. In \mbox{Ref.\ \protect\cite{QDynResp}} the causal Wick theorem for the electromagnetic field was combined with Dyson's standard perturbative approach of quantum field theory \cite{Schweber}. In this paper, we put results of \mbox{Refs.\ \protect\cite{API,APII,APIII,WickCaus,QDynResp}} together. We encounter perfect consistency of generalised phase-space concepts introduced in \mbox{Refs.\ \protect\cite{APII,APIII}} with quantum dynamics in response representation devised in \mbox{Refs.\ \protect\cite{WickCaus,QDynResp}} --- not quite unexpectedly, given that all results are ultimately due to the wave quantisation formula. 

The result of this paper in a nutshell is that, expressed in phase-space terms, dynamics of the electromagnetic field becomes classical. In particular, propagation of the field in space and time is always subject to strict causality. 
Formally, this is due to properties of the free field (ultimately, to Feynman's observation) and to the bilinear structure of the electromagnetic interaction. 

The paper is organised as follows. In section \ref{ch:TNT} we recall conventional \cite{KelleyKleiner,GlauberTN} and amended \cite{APII,APIII} definitions of the time-normal ordering. In section \ref{ch:QR}, we reiterate results of \mbox{Ref.\ \protect\cite{QDynResp}}. 
Sections \ref{ch:G} and \ref{ch:P} are concerned with parallelism between classical stochastic and quantum electrodynamics. In section \ref{ch:G}, we rewrite results of \mbox{Ref.\ \protect\cite{QDynResp}} in terms of time-normally ordered operator averages, and show that this leads to an essentially classical picture of electromagnetic interactions. In section \ref{ch:P}, we introduce the concept of P-functional, which generalises the conventional P-function \cite{MandelWolf} to multitime quantum averages of {Heisenberg}\ operators, and show that any relation for time-normal operator averages and P-functionals coincides with some relation for classical stochastic averages and probability distributions. In section \ref{ch:PS} we demonstrate that P-functionals also give a natural, and in essense classical, insight into the electromagnetic self-action (``dressing'') problem. In section \ref{ch:R} we briefly discuss mathematical complications hidden behind the apparent simplicity of our formulae. The appendix is concerned with functional probability distributions and related issues. 

%********************************************************
\section{Time-normal ordering of operators}%
\label{ch:TNT}
%********************************************************
\subsection{Preliminary remarks}%
\label{ch:TNPR}
%********************************************************
We start from refreshing our memory on the concepts of {\em time-normal operator product and time-normal average\/}. As in \mbox{Ref.\ \protect\cite{QDynResp}} we distinguish the narrow-band\ and broad-band\ case s, which differ in whether the resonance, or rotating wave, approximation (RWA) is or is not made in dynamics. In the narrow-band\ case, definition of the time-normal ordering follows Kelley and Kleiner \cite{KelleyKleiner,GlauberTN}. In the broad-band\ case, we adhere to the amended definition of \mbox{Refs.\ \protect\cite{APII,APIII}}. For formal justifications and discussions see \mbox{Refs.\ \protect\cite{API,APII,APIII,WickCaus,QDynResp}} (cf.\ also \mbox{Refs.\ \protect\cite{Corresp,BWO}}). 

To be specific, we talk about the {Heisenberg}\ dipole-momentum operator $\protect{\hat{\mathcal D}}(t)$ and its Hermitian-adjoint $\protect{\hat{\mathcal D}}^{\dag}(t)$ in the narrow-band\ case, and about the {Heisenberg}\ current operator $\protect{\hat{\mathcal J}}(t)$ in the broad-band\ case. For brevity we drop all arguments of the operators except time. As dynamical quantities, the dipole and current operators will be defined in section \ref{ch:GH}. For purposes of this section, their physical nature is irrelevant. Hermiticity of $\protect{\hat{\mathcal J}}(t)$ does not matter either, with the only exception of reality conditions in section \ref{ch:TNC}. 

%********************************************************
\subsection{The narrow-band\ case}%
\label{ch:TNTR}
%********************************************************
In the narrow-band\ case, time-normal operator ordering is an operation which places all $\protect{\hat{\mathcal D}}^{\dag}(t)$'s to the left of all $\protect{\hat{\mathcal D}}(t)$'s. Among themselves, the $\protect{\hat{\mathcal D}}(t)$ operators are time-ordered, which means setting them from left to right in the order of decreasing time arguments. The $\protect{\hat{\mathcal D}}^{\dag}(t)$ operators are reverse-time-ordered, which means setting them from left to right in the order of increasing time arguments. These two types of time-ordering are denoted as $T_+$ and $T_-$, respectively. That is, 
%=============================================
{\begin{multline}\hspace{0.4\columnwidth}\hspace{-0.4\twocolumnwidth} 
{\mathcal T}{\mbox{\rm\boldmath$:$}}
\protect{\hat{\mathcal D}}^{\dag}(t_1)\cdots 
\protect{\hat{\mathcal D}}^{\dag}(t_m)
\protect{\hat{\mathcal D}}(t_1')\cdots 
\protect{\hat{\mathcal D}}(t_n')
{\mbox{\rm\boldmath$:$}} 
\\ 
= 
T_-
\protect{\hat{\mathcal D}}^{\dag}(t_1)\cdots 
\protect{\hat{\mathcal D}}^{\dag}(t_m)
\, 
T_+
\protect{\hat{\mathcal D}}(t_1')\cdots 
\protect{\hat{\mathcal D}}(t_n') . 
\hspace{0.4\columnwidth}\hspace{-0.4\twocolumnwidth} 
\label{eq:8BS} % \nonumber % \Z 
\end{multline}}%
%+++++++++++++++++++++++++++++++++++++++++++++
The notation ${\mathcal T}{\mbox{\rm\boldmath$:$}}\cdots{\mbox{\rm\boldmath$:$}}$ for the time-normal ordering is borrowed from Mandel and Wolf \cite{MandelWolf}. 

In quantum field theory and condensed matter physics \cite{SchwingerC,Perel,Keldysh}, the double-time-ordered structure as in (\ref{eq:8BS}) is commonly expressed as a closed-time-loop, or C-contour, ordering, which we denote $T_C$. Formally, one marks operators under the $T_{\pm}$-orderings by ${}_{\pm}$ indices, and allows them to commute freely. So, eq.\ (\protect\ref{eq:8BS}) in terms of the $T_C$-ordering becomes, 
%\NBH{: see \Eq{73RQ}}{ 
%=============================================
{\begin{multline}\hspace{0.4\columnwidth}\hspace{-0.4\twocolumnwidth} 
{\mathcal T}{\mbox{\rm\boldmath$:$}}
\protect{\hat{\mathcal D}}^{\dag}(t_1)\cdots 
\protect{\hat{\mathcal D}}^{\dag}(t_m)
\protect{\hat{\mathcal D}}(t_1')\cdots 
\protect{\hat{\mathcal D}}(t_n')
{\mbox{\rm\boldmath$:$}} 
\\ 
= 
T_C
\protect{\hat{\mathcal D}}_-^{\dag}(t_1)\cdots 
\protect{\hat{\mathcal D}}_-^{\dag}(t_m)
\protect{\hat{\mathcal D}}_+(t_1')\cdots 
\protect{\hat{\mathcal D}}_+(t_n') , 
\\ 
= 
T_C
\protect{\hat{\mathcal D}}_+(t_1')\cdots 
\protect{\hat{\mathcal D}}_+(t_n') 
\protect{\hat{\mathcal D}}_-^{\dag}(t_1)\cdots 
\protect{\hat{\mathcal D}}_-^{\dag}(t_m)
. 
\hspace{0.4\columnwidth}\hspace{-0.4\twocolumnwidth} 
\label{eq:73RQ} % \nonumber % \Z 
\end{multline}}%
%+++++++++++++++++++++++++++++++++++++++++++++
etc. For more details see, e.g., our \mbox{Ref.\ \protect\cite{APIII}}. 

%} 
Of actual interest to us are the {\em time-normal averages\/} of the dipole operators, 
%=============================================
{\begin{multline}\hspace{0.4\columnwidth}\hspace{-0.4\twocolumnwidth} 
\protect\protect\ensuremath{\big\langle 
{\mathcal T}{\mbox{\rm\boldmath$:$}}
\protect{\hat{\mathcal D}}^{\dag}(t_1)\cdots 
\protect{\hat{\mathcal D}}^{\dag}(t_m)
\protect{\hat{\mathcal D}}(t_1')\cdots 
\protect{\hat{\mathcal D}}(t_n')
{\mbox{\rm\boldmath$:$}}
\big\rangle} 
\\ 
= \frac{i^m (-i)^n\delta ^{m+n}\Phi\protect\ensuremath{\big(
\nu , \nu^* 
 \big)} 
}{
\delta\nu(t_1)\cdots
\delta\nu(t_m)
\delta\nu^*(t'_1)\cdots
\delta\nu^*(t'_n)
}\settoheight{\auxlv}{$\big|$}%
\raisebox{-0.3\auxlv}{$\big|_{\nu =0}$} . 
\hspace{0.4\columnwidth}\hspace{-0.4\twocolumnwidth} 
\label{eq:11BV} % \nonumber % \Z 
\end{multline}}%
%+++++++++++++++++++++++++++++++++++++++++++++
We have immediately introduced their generating, or characteristic, functional, 
%=============================================
{\begin{multline}\hspace{0.4\columnwidth}\hspace{-0.4\twocolumnwidth} 
\Phi\protect\ensuremath{\big(
\nu , \nu^* 
 \big)} 
= \protect\protect\ensuremath{\bigg\langle 
{\mathcal T}{\mbox{\rm\boldmath$:$}}\exp\protect\protect\ensuremath{\bigg\{
i\int dt\protect\protect\ensuremath{\big[
\nu ^*(t)\protect{\hat{\mathcal D}}(t)-\nu(t)\protect{\hat{\mathcal D}}^{\dag}(t)
\big]} 
\bigg\}} 
{\mbox{\rm\boldmath$:$}}
\bigg\rangle} 
\\ 
= \protect\protect\ensuremath{\bigg\langle 
T_C\exp\protect\protect\ensuremath{\bigg\{
i\int dt\protect\protect\ensuremath{\big[
\nu ^*(t)\protect{\hat{\mathcal D}}_+(t)-\nu(t)\protect{\hat{\mathcal D}}_-^{\dag}(t)
\big]} 
\bigg\}}
\bigg\rangle} 
, 
\hspace{0.4\columnwidth}\hspace{-0.4\twocolumnwidth} 
\label{eq:9BT} % \nonumber % \Z 
\end{multline}}%
%+++++++++++++++++++++++++++++++++++++++++++++
where $\nu (t)$ is an auxiliary complex c-number function. 
The same in terms of the double-time-ordering reads, 
%=============================================
{\begin{multline}\hspace{0.4\columnwidth}\hspace{-0.4\twocolumnwidth} 
\Phi\protect\ensuremath{\big(
\nu , \nu^* 
 \big)} 
= \protect\protect\ensuremath{\bigg\langle 
T_-\exp\protect\protect\ensuremath{\bigg[
-i\int dt\nu(t)\protect{\hat{\mathcal D}}^{\dag}(t)
\bigg]} 
\\ \times 
T_+\exp\protect\protect\ensuremath{\bigg[
i\int dt\nu ^*(t)\protect{\hat{\mathcal D}}(t)
\bigg]} 
\bigg\rangle}
. 
\hspace{0.4\columnwidth}\hspace{-0.4\twocolumnwidth} 
\label{eq:92PM} % \nonumber % \Z 
\end{multline}}%
%+++++++++++++++++++++++++++++++++++++++++++++
The averaging in eqs.\ (\protect\ref{eq:11BV})--(\ref{eq:92PM}) is over the initial ({Heisenberg}) state of the system, 
%=============================================
\protect{\begin{align}{{
 \begin{aligned} 
\protect\protect\ensuremath{\big\langle 
\cdots
\big\rangle} = \text{Tr} \hat \rho (\cdots), 
\end{aligned}}}%
\label{eq:12ZA} % \nonumber % \Z 
\end{align}}%
%+++++++++++++++++++++++++++++++++++++++++++++
where the ellipsis stands for an arbitrary operator. 
Unlike in \mbox{Refs.\ \protect\cite{API,APII,APIII,BWO}}, we define (\ref{eq:9BT}) with a complex-conjugate pair of arguments $\nu(t) , \nu^*(t)$ in place of a pair of independent functions $\nu(t) , \bar\nu(t)$. The reason for this was clarified in \cite{QDynResp}, 
section VB.

A word of extreme caution is in place here. The time-normal ordering (\ref{eq:8BS}), (\ref{eq:73RQ}) is defined only for {\em products\/} of (more generally speaking, for quantities that may be regarded {\em functionals\/} of) $\protect{\hat{\mathcal D}}(t),\protect{\hat{\mathcal D}}^{\dag}(t)$. Ignoring this reservation leads to confusion and plain nonsense. For example, in physical models, dipole operators are commonly defined as, 
%=============================================
\protect{\begin{align}{{
 \begin{aligned} 
\protect{\hat{\mathcal D}}(t) = \hat \psi^{\dag}_{\mathrm{g}}(t) \hat \psi_{\mathrm{e}}(t), 
\end{aligned}}}%
\label{eq:46DJ} % \nonumber % \Z 
\end{align}}%
%+++++++++++++++++++++++++++++++++++++++++++++
where $\hat \psi^{\dag}_{\mathrm{g}}(t)$ and $\hat \psi_{\mathrm{e}}(t)$ are creation and annihilation operators\ for the ground and excited states (say) of an atom. Definitions like (\ref{eq:8BS}), (\ref{eq:73RQ}) may then be given for the atomic operators. To maintain rigour, one has to introduce two symbols, e.g., $\mathcal{T}^{\protect{\hat{\mathcal D}}}{\mbox{\rm\boldmath$:$}}
\cdots
{\mbox{\rm\boldmath$:$}} $ and $\mathcal{T}^{\hat \psi }{\mbox{\rm\boldmath$:$}}
\cdots
{\mbox{\rm\boldmath$:$}} $, for time-normal orderings with respect to dipole momenta and to atomic operators. Then, for the dipole operators, 
%=============================================
{\begin{multline}\hspace{0.4\columnwidth}\hspace{-0.4\twocolumnwidth} 
\mathcal{T}^{\protect{\hat{\mathcal D}}}{\mbox{\rm\boldmath$:$}}\protect{\hat{\mathcal D}}^{\dag}(t)\protect{\hat{\mathcal D}}(t'){\mbox{\rm\boldmath$:$}} = \protect{\hat{\mathcal D}}^{\dag}(t)\protect{\hat{\mathcal D}}(t') 
\\ 
= \hat \psi^{\dag}_{\mathrm{g}}(t) \hat \psi_{\mathrm{e}}(t)
\hat \psi^{\dag}_{\mathrm{g}}(t') \hat \psi_{\mathrm{e}}(t'), 
\hspace{0.4\columnwidth}\hspace{-0.4\twocolumnwidth} 
\label{eq:48DL} % \nonumber % \Z 
\end{multline}}%
%+++++++++++++++++++++++++++++++++++++++++++++
whereas for the atomic operators \cite{endFerm}, 
%=============================================
{\begin{multline}\hspace{0.4\columnwidth}\hspace{-0.4\twocolumnwidth} 
\mathcal{T}^{\hat \psi }{\mbox{\rm\boldmath$:$}}\protect{\hat{\mathcal D}}^{\dag}(t)\protect{\hat{\mathcal D}}(t'){\mbox{\rm\boldmath$:$}} = \mathcal{T}^{\hat \psi }{\mbox{\rm\boldmath$:$}}\hat \psi^{\dag}_{\mathrm{g}}(t) \hat \psi_{\mathrm{e}}(t)
\hat \psi^{\dag}_{\mathrm{g}}(t') \hat \psi_{\mathrm{e}}(t'){\mbox{\rm\boldmath$:$}} \\ 
= 
\bar T\hat \psi^{\dag}_{\mathrm{g}}(t') 
\hat \psi^{\dag}_{\mathrm{g}}(t) 
\, T \hat \psi_{\mathrm{e}}(t)
\hat \psi_{\mathrm{e}}(t') . 
\hspace{0.4\columnwidth}\hspace{-0.4\twocolumnwidth} 
\label{eq:47DK} % \nonumber % \Z 
\end{multline}}%
%+++++++++++++++++++++++++++++++++++++++++++++
These two quantities are distinct for $t\neq t'$ (recall that $\protect{\hat{\mathcal D}}(t),\hat \psi_{\mathrm{g}}(t),\hat \psi_{\mathrm{e}}(t)$ are {Heisenberg}\ operators). By ignoring the difference between $\mathcal{T}^{\protect{\hat{\mathcal D}}}{\mbox{\rm\boldmath$:$}}
\cdots
{\mbox{\rm\boldmath$:$}} $ and $\mathcal{T}^{\hat \psi }{\mbox{\rm\boldmath$:$}}
\cdots
{\mbox{\rm\boldmath$:$}} $ one can ``prove'' that quantities (\ref{eq:48DL}) and (\ref{eq:47DK}) coincide. This is one example of the aforementioned ``plain nonsense.'' Similar reservations apply to other cases of time-normal ordering. 

%********************************************************
\subsection{The broad-band\ case}%
\label{ch:TNTN}
%********************************************************
In the broad-band\ case, the time-normal averages of the current operator are defined through their generating functional, 
%=============================================
\protect{\begin{align}{{
 \begin{aligned} 
\protect\protect\ensuremath{\big\langle 
{\mathcal T}{\mbox{\rm\boldmath$:$}}
\protect{\hat{\mathcal J}}(t_1)\cdots 
\protect{\hat{\mathcal J}}(t_m)
{\mbox{\rm\boldmath$:$}}
\big\rangle} 
%\\ 
= \frac{(-i)^m\delta ^{m}\Phi\protect\ensuremath{\big(
\zeta 
 \big)} 
}{
\delta\zeta (t_1)\cdots
\delta\zeta (t_m)
}\settoheight{\auxlv}{$\big|$}%
\raisebox{-0.3\auxlv}{$\big|_{\zeta =0}$} , 
\end{aligned}}}%
\label{eq:13BX} % \nonumber % \Z 
\end{align}}%
%+++++++++++++++++++++++++++++++++++++++++++++
which is postulated to be \cite{API,APII,APIII}, 
%=============================================
{\begin{multline}\hspace{0.4\columnwidth}\hspace{-0.4\twocolumnwidth} 
\Phi (\zeta ) 
\equiv \protect\protect\ensuremath{\bigg\langle 
{\mathcal T}{\mbox{\rm\boldmath$:$}}\exp \protect\protect\ensuremath{\bigg[
i \int dt \zeta (t) \protect{\hat{\mathcal J}}(t)
\bigg]} {\mbox{\rm\boldmath$:$}}
\bigg\rangle} 
\\ 
= \protect\protect\ensuremath{\bigg\langle 
T_C\exp\protect\protect\ensuremath{\bigg\{
i\int dt\protect\protect\ensuremath{\big[
\zeta^{(-)}(t) \protect{\hat{\mathcal J}}_+(t)
%\\ 
+\zeta ^{(+)}(t) \protect{\hat{\mathcal J}}_-(t)
\big]} 
\bigg\}} 
\bigg\rangle} 
. 
\hspace{0.4\columnwidth}\hspace{-0.4\twocolumnwidth} 
\label{eq:14BY} % \nonumber % \Z 
\end{multline}}%
%+++++++++++++++++++++++++++++++++++++++++++++
The symbols ${}^{(\pm)}$ denote separation of the frequency-positive and negative\ parts of functions, 
%=============================================
\protect{\begin{align}{{
 \begin{aligned} 
& f(t) = f^{(+)}(t) + f^{(-)}(t) , 
\\ 
& f^{(\pm)}(t) = \int_{-\infty}^{+\infty} \frac{d\omega }{2\pi }\mathrm{e}^{-i\omega t}\theta(\pm\omega )
f_{\omega }, & 
& f_{\omega } = \int_{-\infty}^{+\infty} dt \mathrm{e}^{i\omega t}f(t) . 
\end{aligned}}}%
\label{eq:4JH} % \nonumber % \Z 
\end{align}}%
%+++++++++++++++++++++++++++++++++++++++++++++
This operation is alternatively expressed as 
an integral transformation, 
%=============================================
\protect{\begin{align}{{
 \begin{aligned} 
f ^{(\pm)}(t) = \int dt' \delta ^{(\pm)}(t-t') f (t'), 
\end{aligned}}}%
\label{eq:39KV} % \nonumber % \Z 
\end{align}}%
%+++++++++++++++++++++++++++++++++++++++++++++
where 
%=============================================
\protect{\begin{align}{{
 \begin{aligned} 
\delta ^{(\pm)}(t) = \delta ^{(\mp)}(-t) = \protect\protect\ensuremath{\big[
\delta ^{(\mp)}(t)
\big]} ^* = \ensuremath{\pm\frac{1}{2\pi i(t\mp i0^+)}} 
\end{aligned}}}%
\label{eq:40KW} % \nonumber % \Z 
\end{align}}%
%+++++++++++++++++++++++++++++++++++++++++++++
are the frequency-positive and negative\ parts of the delta-function. 
For more details on this operation see \mbox{Ref.\ \protect\cite{APII}}, appendix A. 

Accounting for (\ref{eq:39KV}), eq.\ (\protect\ref{eq:14BY}) reads, 
%=============================================
{\begin{multline}\hspace{0.4\columnwidth}\hspace{-0.4\twocolumnwidth} 
\Phi (\zeta ) %\\ 
= \protect\protect\ensuremath{\bigg\langle 
T_C\exp\protect\protect\ensuremath{\bigg\{
i\int dtdt'\zeta (t) 
\protect\protect\ensuremath{\Big[
\delta ^{(-)}(t-t')\protect{\hat{\mathcal J}}_-(t')
\\ 
+\delta ^{(+)}(t-t')\protect{\hat{\mathcal J}}_+(t')
\Big]} 
\bigg\}} 
\bigg\rangle} . 
\hspace{0.4\columnwidth}\hspace{-0.4\twocolumnwidth} 
\label{eq:20CE} % \nonumber % \Z 
\end{multline}}%
%+++++++++++++++++++++++++++++++++++++++++++++
Differentiating (\ref{eq:20CE}) as per eq.\ (\protect\ref{eq:13BX}) we find the explicit formula, 
%=============================================
{\begin{multline}\hspace{0.4\columnwidth}\hspace{-0.4\twocolumnwidth} 
\protect\protect\ensuremath{\big\langle 
{\mathcal T}{\mbox{\rm\boldmath$:$}}
\protect{\hat{\mathcal J}}(t_1)\cdots 
\protect{\hat{\mathcal J}}(t_m)
{\mbox{\rm\boldmath$:$}}
\big\rangle} 
\\ 
= \protect\protect\ensuremath{\bigg\langle T_C\int dt'_1\cdots dt'_m
\prod_{l=1}^{m}
\protect\protect\ensuremath{\Big[
\delta ^{(-)}(t_l-t'_l)\protect{\hat{\mathcal J}}_-(t'_l)
\\ 
+\delta ^{(+)}(t_l-t'_l)\protect{\hat{\mathcal J}}_+(t'_l)
\Big]} 
\bigg\rangle} . 
\hspace{0.4\columnwidth}\hspace{-0.4\twocolumnwidth} 
\label{eq:21CF} % \nonumber % \Z 
\end{multline}}%
%+++++++++++++++++++++++++++++++++++++++++++++

The GKK definition is recovered applying separation of the frequency-positive and negative\ parts to the operators, 
%=============================================
{\begin{multline}\hspace{0.4\columnwidth}\hspace{-0.4\twocolumnwidth} 
\protect\protect\ensuremath{\big\langle 
{\mathcal T}{\mbox{\rm\boldmath$:$}}
\protect{\hat{\mathcal J}}(t_1)\cdots 
\protect{\hat{\mathcal J}}(t_m)
{\mbox{\rm\boldmath$:$}}
\big\rangle} \settoheight{\auxlv}{$|$}%
\raisebox{-0.3\auxlv}{$|_{\mathrm{GKK}}$}
\\ 
= \protect\protect\ensuremath{\bigg\langle T_C
\prod_{l=1}^{m}
\protect\protect\ensuremath{\big[
\protect{\hat{\mathcal J}}^{(-)}_-(t_l)
%\\ 
+\protect{\hat{\mathcal J}}^{(+)}_+(t_l)
\big]} 
\bigg\rangle}, \ \ \mathrm{(RWA\ only)}
\hspace{0.4\columnwidth}\hspace{-0.4\twocolumnwidth} 
\label{eq:16CA} % \nonumber % \Z 
\end{multline}}%
%+++++++++++++++++++++++++++++++++++++++++++++
which coincides with (\ref{eq:8BS}) up to the replacements, 
%=============================================
\protect{\begin{align}{{
 \begin{aligned} 
& \protect{\hat{\mathcal J}}^{(+)}(t)\leftrightarrow\protect{\hat{\mathcal D}}(t), & 
& \protect{\hat{\mathcal J}}^{(-)}(t)\leftrightarrow\protect{\hat{\mathcal D}}^{\dag}(t). 
\end{aligned}}}%
\label{eq:67JB} % \nonumber % \Z 
\end{align}}%
%+++++++++++++++++++++++++++++++++++++++++++++
In general, eq.\ (\protect\ref{eq:16CA}) is incorrect, because separation of the frequency-positive and negative\ parts in (\ref{eq:21CF}) applies to $T_C$-ordered products of the operators and not to the operators themselves. It becomes a valid approximation under the RWA. For a detailed discussion see \mbox{Refs.\ \protect\cite{APII,QDynResp}}. 
%********************************************************
\subsection{Reality and causality}%
\label{ch:TNC}
%********************************************************
Consistency of all physical interpretations in this paper hinge on reality and causality properties of time-normal products and averages. Using that Hermitian conjugation reverts the order of operators we find, 
%=============================================
{\begin{multline}\hspace{0.4\columnwidth}\hspace{-0.4\twocolumnwidth} 
\protect\protect\ensuremath{\big[
{\mathcal T}{\mbox{\rm\boldmath$:$}}
\hat D^{\dag}(t_1)\cdots 
\hat D^{\dag}(t_m)
\hat D(t_1')\cdots 
\hat D(t_n')
{\mbox{\rm\boldmath$:$}} 
\big]}^{\dag} \\ 
= {\mathcal T}{\mbox{\rm\boldmath$:$}}
\hat D(t_1)\cdots 
\hat D(t_m)
\hat D^{\dag}(t_1')\cdots 
\hat D^{\dag}(t_n')
{\mbox{\rm\boldmath$:$}} . 
\hspace{0.4\columnwidth}\hspace{-0.4\twocolumnwidth} 
\label{eq:45DH} % \nonumber % \Z 
\end{multline}}%
%+++++++++++++++++++++++++++++++++++++++++++++
Adding eq.\ (\protect\ref{eq:40KW}) to the argument and assuming Hermiticity of $\protect{\hat{\mathcal J}}(t)$ it is also straightforward to show that,
%=============================================
\protect{\begin{align}{{
 \begin{aligned} 
\protect\protect\ensuremath{\big[
{\mathcal T}{\mbox{\rm\boldmath$:$}}
\protect{\hat{\mathcal J}}(t_1)\cdots 
\protect{\hat{\mathcal J}}(t_m)
{\mbox{\rm\boldmath$:$}} 
\big]}^{\dag} %\\ 
= {\mathcal T}{\mbox{\rm\boldmath$:$}}
\protect{\hat{\mathcal J}}(t_1)\cdots 
\protect{\hat{\mathcal J}}(t_m)
{\mbox{\rm\boldmath$:$}} . 
\end{aligned}}}%
\label{eq:49DM} % \nonumber % \Z 
\end{align}}% "
%+++++++++++++++++++++++++++++++++++++++++++++
The time-normal averages of the current operator are therefore real, 
%=============================================
\protect{\begin{align}{{
 \begin{aligned} 
\protect\protect\ensuremath{\big\langle 
{\mathcal T}{\mbox{\rm\boldmath$:$}}
\protect{\hat{\mathcal J}}(t_1)\cdots 
\protect{\hat{\mathcal J}}(t_m)
{\mbox{\rm\boldmath$:$}} 
\big\rangle}^* = 
\protect\protect\ensuremath{\big\langle 
{\mathcal T}{\mbox{\rm\boldmath$:$}}
\protect{\hat{\mathcal J}}(t_1)\cdots 
\protect{\hat{\mathcal J}}(t_m)
{\mbox{\rm\boldmath$:$}} 
\big\rangle} , 
\end{aligned}}}%
\label{eq:66EE} % \nonumber % \Z 
\end{align}}%
%+++++++++++++++++++++++++++++++++++++++++++++
while those of the dipole operators obey the natural property, 
%=============================================
{\begin{multline}\hspace{0.4\columnwidth}\hspace{-0.4\twocolumnwidth} 
\protect\protect\ensuremath{\big\langle 
{\mathcal T}{\mbox{\rm\boldmath$:$}}
\hat D^{\dag}(t_1)\cdots 
\hat D^{\dag}(t_m)
\hat D(t_1')\cdots 
\hat D(t_n')
{\mbox{\rm\boldmath$:$}} 
\big\rangle}^* \\ 
= \protect\protect\ensuremath{\big\langle
{\mathcal T}{\mbox{\rm\boldmath$:$}}
\hat D(t_1)\cdots 
\hat D(t_m)
\hat D^{\dag}(t_1')\cdots 
\hat D^{\dag}(t_n')
{\mbox{\rm\boldmath$:$}}
\big\rangle} . 
\hspace{0.4\columnwidth}\hspace{-0.4\twocolumnwidth} 
\label{eq:50DN} % \nonumber % \Z 
\end{multline}}%
%+++++++++++++++++++++++++++++++++++++++++++++

As to causality, the following ``no-peep-into-the-future'' theorem holds: {\em a time-normal product depends on the Heisenberg operators it comprises only for times not later than the latest time argument of these operators\/} \cite{RelCaus}. If dependence of some operators on a perturbation is causal, dependence of their time-normal products on this perturbation is also causal. The question of causality 
of time-normal products reduces to that of quantum equations of motion. The ``no-peep-into-the-future theorem'' extends the causality conditions verified in \cite{APII} from additive external sources in equations of motion to arbitrary perturbations. It holds trivially in the narrow-band\ case, but becomes nontrivial in the broad-band\ case, because separation of the frequency-positive and negative\ parts smears functions all over the time axis. In fact, the ``future tail'' in eq.\ (\protect\ref{eq:21CF}) cancels. For a proof see Ref.\ \cite{RelCaus}. 

\section{Quantum electrodynamics in response representation revisited}%
\label{ch:QR}
%********************************************************
\subsection{The Hamiltonian}%
\label{ch:GH}
%********************************************************
In this section, we reiterate key results of our previous paper \cite{QDynResp}. In \mbox{Ref.\ \protect\cite{QDynResp}}, 
we considered a quantum device interacting with a collection of oscillator modes, with the Hamiltonian in the interaction picture being, 
%=============================================
\protect{\begin{align}{{
 \begin{aligned} 
&\hat H(t) = \hbar 
\sum_{\kappa =1}^{N}\omega_{\kappa}\hat a_{\kappa}^{\dag}\hat a_{\kappa} 
+ \hat H_{\mathrm{dev}}(t) + \hat H_{\text{I}}(t) 
. 
\end{aligned}}}%
\label{eq:86FB} % \nonumber % \Z 
\end{align}}%
%+++++++++++++++++++++++++++++++++++++++++++++
The oscillators, represented by the standard creation and annihilation\ operators, 
%=============================================
\protect{\begin{align}{{
 \begin{aligned} 
& \protect\protect\ensuremath{\big[
\hat a_{\kappa},\hat a_{\kappa '}^{\dag}
\big]} = \delta_{k\kappa '} , & & \kappa ,\kappa'=1,\cdots,N.
\end{aligned}}}%
\label{eq:87FC} % \nonumber % \Z 
\end{align}}%
%+++++++++++++++++++++++++++++++++++++++++++++
are organised in two quantised fields, 
%=============================================
\protect{\begin{align} 
\hat E (x,t) & = i\sum_{\kappa =1}^{M} 
\sqrt{\frac{\hbar\omega_{\kappa}}{2}} u_{\kappa}(x) \hat a_{\kappa} \mathrm{e}^{-i(\omega_{\kappa}-\omega _0)t}, 
\label{eq:4SX} 
\\ 
\hat E^{\dag} (x,t) & = -i\sum_{\kappa =1}^{M} 
\sqrt{\frac{\hbar\omega_{\kappa}}{2}} u^*_{\kappa}(x) \hat a_{\kappa}^{\dag} \mathrm{e}^{i(\omega_{\kappa}-\omega _0)t}, 
\nonumber % \Z 
\\ 
\hat A(x,t)& = \sum_{\kappa =M+1}^{N}
\sqrt{\frac{\hbar}{2\omega_{\kappa}}}
 u_{\kappa}(x) \hat a_{\kappa}\mathrm{e}^{-i\omega_{\kappa}t} + \mathrm{H.c.}\, , 
\label{eq:45LB} % \nonumber % \Z 
\end{align}}%
%+++++++++++++++++++++++++++++++++++++++++++++
where $ u_k(x)$ are complex mode functions, and variable $x$ comprises all field arguments except time. Electromagnetic interaction in (\ref{eq:86FB}) is split accordingly, 
%=============================================
\protect{\begin{align}{{
 \begin{aligned} 
\hat H_{\text{I}}(t) = \hat H_{\text{I}} ^{(1)}(t) + \hat H_{\text{I}} ^{(2)}(t) . 
\end{aligned}}}%
\label{eq:82JT} % \nonumber % \Z 
\end{align}}%
%+++++++++++++++++++++++++++++++++++++++++++++
The {\em narrow-band\/}, or {\em resonant\/}, field $\hat E(x,t)$ interacts with the device according to the resonant\ Hamiltonian, 
%=============================================
{\begin{multline}\hspace{0.4\columnwidth}\hspace{-0.4\twocolumnwidth} 
\hat H_{\text{I}} ^{(1)}(t) = 
- \int dx\protect\protect\ensuremath{\big[ 
\hat D (x,t)\hat E ^{\dag}(x,t) + \hat D (x,t)E^*_{\mathrm{e}}(x,t) 
\\ 
+ D_{\mathrm{e}}(x,t)\hat E ^{\dag}(x,t)
\big]}+\mathrm{H.c.}\, , 
\hspace{0.4\columnwidth}\hspace{-0.4\twocolumnwidth} 
\label{eq:83AR} 
\end{multline}}%
%+++++++++++++++++++++++++++++++++++++++++++++
while the {\em broad-band\/}, or {\em nonresonant\/}, field $\hat A(x,t)$ --- 
according to the nonresonant\ Hamiltonian, 
%=============================================
{\begin{multline}\hspace{0.4\columnwidth}\hspace{-0.4\twocolumnwidth} 
\hat H_{\text{I}}^{(2)}(t) = 
- 
\int dx \protect\protect\ensuremath{\big[
\hat J(x,t)\hat A(x,t)+\hat J(x,t)A_{\mathrm{e}}(x,t)
\\ 
+ J_{\mathrm{e}}(x,t)\hat A(x,t)
\big]} 
 . 
\hspace{0.4\columnwidth}\hspace{-0.4\twocolumnwidth} 
\label{eq:81AP} % \nonumber % \Z 
\end{multline}}%
%+++++++++++++++++++++++++++++++++++++++++++++
The Hamiltonian $\hat H_{\mathrm{dev}}(t)$, the dipole momentum $\hat D (x,t)$ and the current operator $\hat J(x,t)$ describe the device. They commute with $\hat a_{\kappa },\hat a_{\kappa }^{\dag}$ and otherwise remain arbitrary. The initial state of all oscillators is vacuum, while that of the device is also arbitrary. The c-number external sources $E_{\mathrm{e}}(x,t)$, $D_{\mathrm{e}}(x,t)$, $ A_{\mathrm{e}}(x,t)$, and $ J_{\mathrm{e}}(x,t)$ are added for formal purposes. 

Hamiltonian (\ref{eq:86FB}) may be adjusted to any conceivable case of electromagnetic interaction. From our perspective, this Hamiltonian is a structural model of a quantum-optical experiment involving photodetection. For justification and discussion of this model see 
sections II and III in \mbox{Ref.\ \protect\cite{QDynResp}}. 

%********************************************************
\subsection{Condensed notation}%
\label{ch:DN}
%********************************************************
To keep the bulk of formulae under the lid and make their structure more transparent, we make extensive use of condensed notation, 
%=============================================
\protect{\begin{align} 
& fg = \int dx dt f(x,t) g(x,t) , 
\label{eq:3VS} % \nonumber % \Z 
\\
& fKg = \int dx dx' dt dt' f(x,t)K(x,x',t-t') 
%\\ \times 
g(x',t') , 
\label{eq:76NU} % \nonumber % \Z 
\\ 
& (Kg)(x,t) = \int dx' dt' K(x,x',t-t')g(x',t') , 
\label{eq:77NV} % \nonumber % \Z 
\\ 
& (fK)(x,t) = \int dx' dt' g(x',t')K(x',x,t'-t) , 
\label{eq:78NW} % \nonumber % \Z 
\end{align}}%
%+++++++++++++++++++++++++++++++++++++++++++++
where $f(x,t)$ and $g(x,t)$ are c-number or q-number functions, and $K(x,x',t-t')$ is a c-number kernel. The ``products'' $fg$ and $fKg$ denote scalars, while $Kg$ and $fK$ --- functions (fields). 

%********************************************************
\subsection{Closed-time-loop formalism and response transformation}%
\label{ch:UD}
%********************************************************
Fields, currents and dipoles in eqs.\ (\protect\ref{eq:4SX})--(\ref{eq:81AP}) are interaction-picture operators. Their {Heisenberg}\ counterparts will be denoted by calligraphic letters as $\protect{\hat{\mathcal E}}(x,t)$, $\protect{\hat{\mathcal A}}(x,t)$, $\protect{\hat{\mathcal D}}(x,t)$, and $\protect{\hat{\mathcal J}}(x,t)$. We solve for the characteristic functional of the $T_C$-ordered products of the Heisenberg operators written in {\em causal variables\/}, 
\begin{widetext} 
%=============================================
{\begin{multline} 
\Phi\protect\ensuremath{\big(
\eta ,
\mu ,\mu ^*, 
\zeta ,
\nu ,\nu ^*
\big | 
j_{\mathrm{e}},
d_{\mathrm{e}},d_{\mathrm{e}}^*, 
a_{\mathrm{e}},
e_{\mathrm{e}},e_{\mathrm{e}}^*
\big | 
J_{\mathrm{e}},
D_{\mathrm{e}},D_{\mathrm{e}}^*,
A_{\mathrm{e}},
E_{\mathrm{e}},E_{\mathrm{e}}^*
 \big)} 
\\ 
= \protect\protect\ensuremath{\Big\langle T_C 
\exp\protect\ensuremath{\big(
i{\eta}_+ \protect{\hat{\mathcal A}}_+ 
-i{\eta}_- \protect{\hat{\mathcal A}}_- 
+i{\zeta}_+ \protect{\hat{\mathcal J}}_+ 
-i{\zeta}_- \protect{\hat{\mathcal J}}_-
 \big)} 
\\ \times 
\exp\protect\ensuremath{\big(
i\bar\mu_+ \protect{\hat{\mathcal E}}_+ +i\mu_+ \protect{\hat{\mathcal E}}_+^{\dag} 
-i\bar\mu_- \protect{\hat{\mathcal E}}_- -i\mu_- \protect{\hat{\mathcal E}}_-^{\dag}
+i\bar\nu_+ \protect{\hat{\mathcal D}}_+ +i\nu_+ \protect{\hat{\mathcal D}}_+^{\dag} 
-i\bar\nu_- \protect{\hat{\mathcal D}}_- -i\nu_- \protect{\hat{\mathcal D}}_-^{\dag}
 \big)} 
\Big\rangle}\settoheight{\auxlv}{$\big|$}%
\raisebox{-0.3\auxlv}{$\big|_{\mathrm{c.v.}}$} , 
\label{eq:93FK} % \nonumber % \Z 
\end{multline}}%
%+++++++++++++++++++++++++++++++++++++++++++++
where $\eta_{\pm}(x,t)$, 
$\zeta_{\pm}(x,t)$, 
$\mu_{\pm}(x,t)$, 
$\bar\mu_{\pm}(x,t)$, 
$\nu_{\pm}(x,t)$, and 
$\bar\nu_{\pm}(x,t)$ in (\ref{eq:93FK}) are auxiliary complex c-number functions, and 
c.v.\ refers to the set of {\em response substitutions\/}, (with arguments dropped for clarity)
%=============================================
\protect{\begin{align}{{
 \begin{aligned} 
\eta_{\pm}
%(x,t) 
& = \frac{ j_{\mathrm{e}}
%(x,t) 
}{\hbar }\pm \eta ^{(\mp)}
%(x,t) 
, 
&
\zeta_{\pm}
%(x,t) 
 & 
= \frac{ a_{\mathrm{e}}
%(x,t) 
}{\hbar }\pm \zeta ^{(\mp)}
%(x,t) 
, 
\\ 
\mu_+
%(x,t) 
 & = \frac{d_{\mathrm{e}}
%(x,t) 
}{\hbar }, & 
\bar\mu_+
%(x,t) 
 & = \mu^*
%(x,t) 
+ \frac{d_{\mathrm{e}}^*
%(x,t) 
}{\hbar }, & 
\bar\mu_-
%(x,t) 
 & = \frac{d_{\mathrm{e}}^*
%(x,t) 
}{\hbar }, & 
\mu_-
%(x,t) 
 & = \mu
%(x,t) 
 + \frac{d_{\mathrm{e}}
%(x,t) 
}{\hbar }, 
\\ 
\nu_+
%(x,t) 
 & = \frac{e_{\mathrm{e}}
%(x,t) 
}{\hbar }, & 
\bar\nu_+
%(x,t) 
 & = \nu^*
%(x,t) 
+ \frac{e_{\mathrm{e}}^*
%(x,t) 
}{\hbar }, & 
\bar\nu_-
%(x,t) 
 & = \frac{e_{\mathrm{e}}^*
%(x,t) 
}{\hbar }, & 
\nu_-
%(x,t) 
 & = \nu
%(x,t) 
 + \frac{e_{\mathrm{e}}
%(x,t) 
}{\hbar }. 
\end{aligned}}}%
\label{eq:21FB} % \nonumber % \Z 
\end{align}}%
%+++++++++++++++++++++++++++++++++++++++++++++
\end{widetext}%
We use notation (\ref{eq:3VS}). 
The symbols ${}^{(\pm)}$ denote separation of the frequency-positive and negative\ parts, cf.\ eq.\ (\protect\ref{eq:4JH}). $T_C$-ordering was defined in section \ref{ch:TNTR}. 
The averaging in (\ref{eq:93FK}) is over the initial (Heisenberg) state of the system, cf.\ eq.\ (\protect\ref{eq:12ZA}). We remind that the initial state of all oscillators is vacuum, 
%=============================================
\protect{\begin{align}{{
 \begin{aligned} 
\hat \rho = \hat \rho_{\mathrm{dev}} \otimes 
\protect\protect\ensuremath{\big| 
0
\big\rangle } 
\protect\protect\ensuremath{\big\langle 
0
\big|} , 
\end{aligned}}}%
\label{eq:13ZB} % \nonumber % \Z 
\end{align}}%
%+++++++++++++++++++++++++++++++++++++++++++++

Not to be lost in these definitions, note the following. The {Heisenberg}\ operators are by construction dependent (conditional) on the sources; in (\ref{eq:93FK}), this dependence is made explicit. Furthermore, operators in (\ref{eq:93FK}) are organised in repetitive structures. So, the current operator $\protect{\hat{\mathcal J}}(x,t)$ and variables $\zeta(x,t) $, $a_{\textrm{e}}(x,t)$ emerge as a combination, 
%=============================================
{\begin{multline} 
i\zeta ^{(-)} \protect{\hat{\mathcal J}}_+
+ 
i\zeta ^{(+)} \protect{\hat{\mathcal J}}_- 
+ \frac{ia_{\textrm{e}}\protect\ensuremath{\big(
\protect{\hat{\mathcal J}}_+ - \protect{\hat{\mathcal J}}_-
 \big)}}{\hbar } \\ 
= i\int dx dt \protect\protect\ensuremath{\bigg\{
\zeta ^{(-)}(x,t) \protect{\hat{\mathcal J}}_+(x,t)
+ 
\zeta ^{(+)}(x,t) \protect{\hat{\mathcal J}}_-(x,t) 
\\ 
+ \frac{a_{\textrm{e}}(x,t)\protect\protect\ensuremath{\big[
\protect{\hat{\mathcal J}}_+(x,t) - \protect{\hat{\mathcal J}}_-(x,t)
\big]}}{\hbar }
\bigg\}} . 
\label{eq:59HT} % \nonumber % \Z 
\end{multline}}%
%+++++++++++++++++++++++++++++++++++++++++++++
To better orient the reader, we have also expanded the condensed notation. The nonresonant\ part of the field $\protect{\hat{\mathcal A}}(x,t)$ and variables $\eta(x,t) $, $j_{\textrm{e}}(x,t)$ are organised in a similar combination. The dipole operator $\protect{\hat{\mathcal D}}(x,t)$ and variables $\nu(x,t) $, $e_{\textrm{e}}(x,t)$ enter as a structure, 
%=============================================
\protect{\begin{align}{{
 \begin{aligned} 
i\nu ^*\protect{\hat{\mathcal D}}_+
-i\nu\protect{\hat{\mathcal D}}^{\dag}_- 
+ \frac{ie_{\textrm{e}}^*\protect\ensuremath{\big(
\protect{\hat{\mathcal D}}_+-\protect{\hat{\mathcal D}}_-
 \big)} +ie_{\textrm{e}}\protect\ensuremath{\big(
\protect{\hat{\mathcal D}}^{\dag}_+-\protect{\hat{\mathcal D}}^{\dag}_-
 \big)} }{\hbar } . 
\end{aligned}}}%
\label{eq:60HU} % \nonumber % \Z 
\end{align}}%
%+++++++++++++++++++++++++++++++++++++++++++++
The resonant\ part of the field $\protect{\hat{\mathcal E}}(x,t)$ and variables $\mu(x,t) $, $d_{\textrm{e}}(x,t)$ are organised similarly. Formal patterns characteristic of the narrow-band\ case\ are in fact a resonance approximation to those characteristic of the broad-band\ case, cf.\ \mbox{Ref.\ \protect\cite{QDynResp}}, 
appendix B. 

%********************************************************
\subsection{Consistency conditions}%
\label{ch:UPC}
%********************************************************
The critical property of functional (\ref{eq:93FK}) is that it depends only on sums of the external sources $A_{\mathrm{e}}$, $J_{\mathrm{e}}$, $E_{\mathrm{e}}$, $D_{\mathrm{e}}$ and the corresponding auxiliary variables $a_{\mathrm{e}}$, $j_{\mathrm{e}}$, $e_{\mathrm{e}}$, $d_{\mathrm{e}}$ \cite{QDynResp}: 
\begin{widetext} 
%=============================================
{\begin{multline} 
\Phi\protect\ensuremath{\big(
\eta ,\mu ,\mu ^*,\zeta, \nu ,\nu ^*
\big | 
j_{\mathrm{e}},
d_{\mathrm{e}},d_{\mathrm{e}}^*, 
a_{\mathrm{e}},
e_{\mathrm{e}},e_{\mathrm{e}}^*
\big | 
J_{\mathrm{e}},
D_{\mathrm{e}},D_{\mathrm{e}}^*,
A_{\mathrm{e}},
E_{\mathrm{e}},E_{\mathrm{e}}^*
 \big)} \\ 
= 
\Phi\protect\ensuremath{\big(
\eta ,\mu ,\mu ^*,\zeta, \nu ,\nu ^*
\big | 
0,0,0,
0,0,0 
\big | 
J_{\mathrm{e}}+j_{\mathrm{e}},
D_{\mathrm{e}}+d_{\mathrm{e}},D_{\mathrm{e}}^*+d_{\mathrm{e}}^*,
A_{\mathrm{e}}+a_{\mathrm{e}},
E_{\mathrm{e}}+e_{\mathrm{e}},E_{\mathrm{e}}^*+e_{\mathrm{e}}^*
 \big)} 
. 
\label{eq:29FL} % \nonumber % \Z 
\end{multline}}%
%+++++++++++++++++++++++++++++++++++++++++++++
%\end{widetext}%
Alternatively, 
%=============================================
{\begin{multline} 
\Phi\protect\ensuremath{\big(
\eta ,\mu ,\mu ^*,\zeta, \nu ,\nu ^*
\big | 
j_{\mathrm{e}},
d_{\mathrm{e}},d_{\mathrm{e}}^*, 
a_{\mathrm{e}},
e_{\mathrm{e}},e_{\mathrm{e}}^*
\big | 
J_{\mathrm{e}},
D_{\mathrm{e}},D_{\mathrm{e}}^*,
A_{\mathrm{e}},
E_{\mathrm{e}},E_{\mathrm{e}}^*
 \big)} 
\\ 
= 
\Phi\protect\ensuremath{\big(
\eta ,\mu ,\mu ^*,\zeta, \nu ,\nu ^*
\big | 
J_{\mathrm{e}}+j_{\mathrm{e}},
D_{\mathrm{e}}+d_{\mathrm{e}},D_{\mathrm{e}}^*+d_{\mathrm{e}}^*,
A_{\mathrm{e}}+a_{\mathrm{e}},
E_{\mathrm{e}}+e_{\mathrm{e}},E_{\mathrm{e}}^*+e_{\mathrm{e}}^*
\big | 
0,0,0,
0,0,0 
 \big)} 
. 
\label{eq:38FV} % \nonumber % \Z 
\end{multline}}%
%+++++++++++++++++++++++++++++++++++++++++++++
These relations are a generalisation of {\em consistency conditions\/} derived in \mbox{Refs.\ \protect\cite{APII,APIII}}. Equations (\ref{eq:29FL}) and (\ref{eq:38FV}) express the same functional but much differ in their interpretation. Equation (\ref{eq:38FV}) shows that, mathematically, external c-number sources in quantum equations of motion are redundant. Information contained in the {Heisenberg}\ operators conditional on the sources is already present in the operators defined without the sources. This is an important fact, because c-number sources are formal and, strictly speaking, unphysical quantities. At the same time, quantum system evolving under the influence of external sources is a very convenient formal viewpoint; in many cases, it is also a valid macroscopic approximation. This {\em response viewpoint\/}, expressed by eq.\ (\protect\ref{eq:29FL}), is the one we adhere to in this paper. 

In view of eq.\ (\protect\ref{eq:29FL}) we may set the redundant auxiliary variables to zero, 
%=============================================
\protect{\begin{align}{{
 \begin{aligned} 
a_{\mathrm{e}}(x,t) = j_{\mathrm{e}}(x,t) = e_{\mathrm{e}}(x,t) = d_{\mathrm{e}}(x,t) = 0. 
\end{aligned}}}%
\label{eq:24FE} % \nonumber % \Z 
\end{align}}%
%+++++++++++++++++++++++++++++++++++++++++++++
Formal description of the system is then given by the reduced characteristic functional, 
%=============================================
\protect{\begin{align}{{
 \begin{aligned} 
\Phi\protect\ensuremath{\big(
\eta ,\mu ,\mu ^*,\zeta, \nu ,\nu ^*
\big | 
J_{\mathrm{e}},
D_{\mathrm{e}},D_{\mathrm{e}}^*,
A_{\mathrm{e}},
E_{\mathrm{e}},E_{\mathrm{e}}^* 
 \big)} 
%\\ 
= 
\Phi\protect\ensuremath{\big(
\eta ,\mu ,\mu ^*,\zeta, \nu ,\nu ^*
\big | 
0,0,0,
0,0,0 
\big | 
J_{\mathrm{e}},
D_{\mathrm{e}},D_{\mathrm{e}}^*,
A_{\mathrm{e}},
E_{\mathrm{e}},E_{\mathrm{e}}^*
 \big)} 
. 
\end{aligned}}}%
\label{eq:77JN} % \nonumber % \Z 
\end{align}}%
%+++++++++++++++++++++++++++++++++++++++++++++
\end{widetext}%
This does not lead to any loss of generality. Full quantum formulae may be recovered replacing, 
%=============================================
\protect{\begin{align}{{
 \begin{aligned} 
& A_{\mathrm{e}}\to a_{\mathrm{e}}+A_{\mathrm{e}}, & 
& J_{\mathrm{e}}\to j_{\mathrm{e}}+J_{\mathrm{e}}, \\ 
& E_{\mathrm{e}}\to e_{\mathrm{e}}+E_{\mathrm{e}}, & 
& D_{\mathrm{e}}\to d_{\mathrm{e}}+D_{\mathrm{e}} . 
\end{aligned}}}%
\label{eq:31FN} % \nonumber % \Z 
\end{align}}%
%+++++++++++++++++++++++++++++++++++++++++++++

%********************************************************
\subsection{Reduction to currents and dipoles}%
\label{ch:UPD}
%********************************************************
Full electromagnetic properties of the quantum device may be expressed by the properties of the {Heisenberg}\ (``dressed'') current and dipole operators. They are contained in the functional, 
%=============================================
{\begin{multline}\hspace{0.4\columnwidth}\hspace{-0.4\twocolumnwidth} 
\Phi_{\mathrm{dev}}\protect\ensuremath{\big(
\zeta ,\nu ,\nu ^*
\big | 
A_{\mathrm{e}},
E_{\mathrm{e}},E_{\mathrm{e}}^* 
 \big)} \\ 
= \Phi\protect\ensuremath{\big(
0 ,0,0,\zeta, \nu ,\nu ^*
\big | 
0,
0,0,
A_{\mathrm{e}},
E_{\mathrm{e}},E_{\mathrm{e}}^* 
 \big)} . 
\hspace{0.4\columnwidth}\hspace{-0.4\twocolumnwidth} 
\label{eq:90FF} % \nonumber % \Z 
\end{multline}}%
%+++++++++++++++++++++++++++++++++++++++++++++
A formula reducing (\ref{eq:93FK}) to (\ref{eq:90FF}) was found in \mbox{Ref.\ \protect\cite{QDynResp}}. Under conditions (\ref{eq:24FE}) it reads, 
%=============================================
{\begin{multline} 
\Phi\protect\ensuremath{\big(
\eta ,\mu ,\mu ^*,\zeta , \nu ,\nu ^*
\big | 
J_{\mathrm{e}},
D_{\mathrm{e}},D_{\mathrm{e}}^*,
A_{\mathrm{e}},
E_{\mathrm{e}},E_{\mathrm{e}}^* 
 \big)} \\ 
= \exp\protect\ensuremath{\big(
i\eta\Delta_{\text{R}} J_{\mathrm{e}} 
+i\mu ^*G_{\text{R}} D_{\mathrm{e}}
-i\mu G_{\text{R}} ^* D_{\mathrm{e}} ^*
 \big)} 
\\ \times 
\Phi_{\mathrm{dev}}\protect\ensuremath{\big(
\zeta+\eta\Delta_{\text{R}} ,\nu+\mu G_{\text{R}}^* ,\nu^*+\mu^* G_{\text{R}} 
\big | 
A_{\mathrm{ext}},E_{\mathrm{ext}},E_{\mathrm{ext}}^*
 \big)} . 
\label{eq:54XX} % \nonumber % \Z 
\end{multline}}%
%+++++++++++++++++++++++++++++++++++++++++++++
We use here abbreviated notation (\ref{eq:76NU}) and (\ref{eq:78NW}). 
The kernels $\Delta_{\text{R}}$ and $G_{\text{R}}$ given by the formulae, 
%=============================================
\protect{\begin{align} 
&\Delta_{\text{R}}(x,x',t-t') = \frac{i}{\hbar }\theta(t-t') 
\protect\protect\ensuremath{\big[
\hat A(x,t),\hat A(x',t')
\big]} 
, 
\label{eq:83LF} % \nonumber % \Z 
\\ 
& G_{\text{R}}(x,x',t-t') = \frac{i}{\hbar}\theta(t-t')
\protect\protect\ensuremath{\big[
\hat E(x,t),\hat E^{\dag}(x',t')
\big]} 
, 
\label{eq:85LJ} % \nonumber % \Z 
\end{align}}%
%+++++++++++++++++++++++++++++++++++++++++++++
and the external fields are combinations of the sources, 
%=============================================
\protect{\begin{align}{{
 \begin{aligned} 
 A_{\mathrm{ext}}(x,t) & = A_{\mathrm{e}}(x,t) 
+ \int dx'dt' \Delta_{\text{R}}(x,x',t-t') J_{\mathrm{e}}(x',t')
, \\ 
E_{\mathrm{ext}}(x,t) & = E_{\mathrm{e}}(x,t) 
+ \int dx'dt' G_{\text{R}}(x,x',t-t') D_{\mathrm{e}}(x',t')
. 
\end{aligned}}}%
\label{eq:26FH} % \nonumber % \Z 
\end{align}}%
%+++++++++++++++++++++++++++++++++++++++++++++
% hide 
Definitions (\ref{eq:83LF}), (\ref{eq:85LJ}) are Kubo's formulae for linear response functions \cite{Kubo}; for more details see \mbox{Ref.\ \protect\cite{API}}. Commutators in (\ref{eq:85LJ}) and (\ref{eq:83LF}) are c-numbers so that quantum averaging present in Kubo's formula is dropped. In other words, response of a linear system does not depend on its state. Explicit expressions for $G_{\text{R}}$ and $\Delta_{\text{R}}$ are found from definitions (\ref{eq:4SX}) and (\ref{eq:45LB}), see \mbox{Ref.\ \protect\cite{QDynResp}}. 

%********************************************************
\subsection{``Dressing'' of currents and dipoles}%
\label{ch:UPS}
%********************************************************
Nontrivial part of perturbative calculations is formally expressed by the {\em dressing formula\/} \cite{QDynResp},
%=============================================
{\begin{multline}\hspace{0.4\columnwidth}\hspace{-0.4\twocolumnwidth} 
\Phi_{\mathrm{dev}}\protect\ensuremath{\big(
\zeta ,\nu,\nu ^*\big|a_{\mathrm{e}},e_{\mathrm{e}},e_{\mathrm{e}}^*
 \big)} \\ 
= 
\exp\protect\ensuremath{\bigg(
-i\frac{\delta }{\delta a_{\mathrm{e}}}\Delta_{\text{R}}
\frac{\delta }{\delta \zeta }
-i\frac{\delta }{\delta e_{\mathrm{e}}}G_{\text{R}}
\frac{\delta }{\delta \nu^* }
+i\frac{\delta }{\delta e_{\mathrm{e}}^*}G_{\text{R}}^*
\frac{\delta }{\delta \nu }
 \bigg)}
\\ \times 
\Phi_{\mathrm{dev}}^{\mathrm{I}}\protect\ensuremath{\big(
\zeta ,\nu,\nu ^* \big| a_{\mathrm{e}},e_{\mathrm{e}},e_{\mathrm{e}}^*
 \big)} , 
\hspace{0.4\columnwidth}\hspace{-0.4\twocolumnwidth} 
\label{eq:91FH} % \nonumber % \Z 
\end{multline}}%
%+++++++++++++++++++++++++++++++++++++++++++++
where $\Phi_{\mathrm{dev}}^{\mathrm{I}}$ contains properties of the interaction-picture (``bare'') current and dipole operators, 
%=============================================
{\begin{multline}\hspace{0.4\columnwidth}\hspace{-0.4\twocolumnwidth} 
\Phi_{\mathrm{dev}}^{\mathrm{I}}\protect\ensuremath{\big(
\zeta ,\nu ,\nu ^*\big| a_{\mathrm{e}},e_{\mathrm{e}},e_{\mathrm{e}}^*
 \big)} 
= \text{Tr}\hat\rho_{\mathrm{dev}}
T_C\exp\protect\ensuremath{\big(
i{\zeta}_+ \hat J_{+} 
-i{\zeta}_- \hat J_{-}
\\ 
+i\bar\nu_+ \hat D_{+} +i\nu_+ \hat D_{+}^{\dag}
-i\bar\nu_- \hat D_{-} -i\nu_- \hat D_{-}^{\dag}
 \big)} 
\settoheight{\auxlv}{$\big|$}%
\raisebox{-0.3\auxlv}{$\big|_{\mathrm{c.v.}}$} , 
\hspace{0.4\columnwidth}\hspace{-0.4\twocolumnwidth} 
\label{eq:92FJ} % \nonumber % \Z 
\end{multline}}%
%+++++++++++++++++++++++++++++++++++++++++++++
and c.v.\ refers to a suitable subset of eqs.\ (\protect\ref{eq:21FB}). 

Of importance for consistency of all our interpretations is that $\Phi_{\mathrm{dev}}^{\mathrm{I}}$ may equally be written in a response form, 
%=============================================
{\begin{multline}\hspace{0.4\columnwidth}\hspace{-0.4\twocolumnwidth} 
\Phi_{\mathrm{dev}}^{\mathrm{I}}\protect\ensuremath{\big(
\zeta ,\nu ,\nu ^*\big| A_{\mathrm{e}},E_{\mathrm{e}},E_{\mathrm{e}}^*
 \big)} \\ 
= \text{Tr}\hat\rho_{\mathrm{dev}}
{\mathcal T}{\mbox{\rm\boldmath$:$}}\exp\protect\ensuremath{\big(
i{\zeta} \hat J' 
+i\bar\nu^* \hat D' -i\nu \hat D^{\prime\dag}
 \big)} 
{\mbox{\rm\boldmath$:$}} 
, 
\hspace{0.4\columnwidth}\hspace{-0.4\twocolumnwidth} 
\label{eq:64HY} % \nonumber % \Z 
\end{multline}}%
%+++++++++++++++++++++++++++++++++++++++++++++
where the primed operators are defined as {Heisenberg}\ ones with respect to the Hailtonian, 
%=============================================
{\begin{multline}\hspace{0.4\columnwidth}\hspace{-0.4\twocolumnwidth} 
\hat H'(t) = \hat H_{\textrm{dev}}(t) - \int dx \protect\protect\ensuremath{\big[
A_{\textrm{e}}(x,t)\hat J(x,t)
\\ 
+E^*_{\textrm{e}}(x,t)\hat D(x,t)
+E_{\textrm{e}}(x,t)\hat D^{\dag}(x,t)
\big]} . 
\hspace{0.4\columnwidth}\hspace{-0.4\twocolumnwidth} 
\label{eq:63HX} % \nonumber % \Z 
\end{multline}}%
%+++++++++++++++++++++++++++++++++++++++++++++
This is Hamiltonian (\ref{eq:86FB}) with field operators set to zero. Consequently, equivalence of definitions (\ref{eq:92FJ}) and (\ref{eq:64HY}) is a particular case of consistency condition (\ref{eq:29FL}), with all arguments related to fields dropped. 

%********************************************************
\section{Conditional time-normal averages}%
\label{ch:G}
%********************************************************
\subsection{Characteristic functionals as time-normal averages}%
\label{ch:GF}
%********************************************************
An astonishing feature of eqs.\ (\protect\ref{eq:54XX}) and (\ref{eq:91FH}) is that they lack Planck's constant. These equations provide an exact, albeit formal, solution to the problem of electromagnetic interaction in quantum mechanics. Planck's constant is present in the definition of the fields (\ref{eq:4SX}), (\ref{eq:45LB}) and of the response functions (\ref{eq:85LJ}), (\ref{eq:83LF}), and in the response substitutions (\ref{eq:21FB}), but falls out of the final formulae. Equations (\ref{eq:54XX}) and (\ref{eq:91FH}) survive the classical limit $\hbar \to 0$ without changes, and must therefore exist in classical statistical electrodynamics. 

The most natural correspondence between quantum and classical electrodynamics emerges if we rewrite the key equation (\ref{eq:54XX}) in terms of the time-normal averages introduced in section \ref{ch:TNT}. Taking notice of eqs.\ (\protect\ref{eq:59HT}), (\ref{eq:60HU}) we find the explicit q-number formula for the reduced functional (\ref{eq:77JN}), 
%=============================================
{\begin{multline} 
\Phi\protect\ensuremath{\big(
\eta ,\mu ,\mu ^*, \zeta ,\nu ,\nu ^*
\big | 
J_{\mathrm{e}},
D_{\mathrm{e}},D_{\mathrm{e}}^*,
A_{\mathrm{e}},
E_{\mathrm{e}},E_{\mathrm{e}}^* 
 \big)} \\ 
= \protect\protect\ensuremath{\Big\langle 
T_C \exp\protect\protect\ensuremath{\big[
i\eta ^{(-)} \protect{\hat{\mathcal A}}_+
+i\eta ^{(+)} \protect{\hat{\mathcal A}}_-
+i\zeta ^{(-)} \protect{\hat{\mathcal J}}_+
+i\zeta ^{(+)} \protect{\hat{\mathcal J}}_-
\\ 
+i\mu^* \protect{\hat{\mathcal E}}_+
-i\mu \protect{\hat{\mathcal E}}^{\dag}_-
+i\nu^* \protect{\hat{\mathcal D}}_+
-i\nu \protect{\hat{\mathcal D}}^{\dag}_-
\big]} 
\Big\rangle} . 
\label{eq:66JA} % \nonumber % \Z 
\end{multline}}%
%+++++++++++++++++++++++++++++++++++++++++++++
We remind that the Heisenberg operators are by construction conditional on the external sources in Hamiltonian (\ref{eq:86FB}). 
Comparing (\ref{eq:66JA}) to eqs.\ (\protect\ref{eq:9BT}) and (\ref{eq:14BY}) we find, 
%=============================================
{\begin{multline} 
\Phi\protect\ensuremath{\big(
\eta ,\mu ,\mu ^*, \zeta ,\nu ,\nu ^*
\big | 
J_{\mathrm{e}},
D_{\mathrm{e}},D_{\mathrm{e}}^*,
A_{\mathrm{e}},
E_{\mathrm{e}},E_{\mathrm{e}}^* 
 \big)} 
\\ 
= 
\protect\protect\ensuremath{\Big\langle {\mathcal T}{\mbox{\rm\boldmath$:$}}\exp\protect\ensuremath{\big(
i\eta\protect{\hat{\mathcal A}}+i\zeta\protect{\hat{\mathcal J}}+i\mu^*\protect{\hat{\mathcal E}}-i\mu\protect{\hat{\mathcal E}}^{\dag}
+i\nu^* \protect{\hat{\mathcal D}}-i\nu\protect{\hat{\mathcal D}}^{\dag}
 \big)} 
{\mbox{\rm\boldmath$:$}}
\Big\rangle} . 
\label{eq:61HV} % \nonumber % \Z 
\end{multline}}%
%+++++++++++++++++++++++++++++++++++++++++++++
That is, {\em under conditions (\ref{eq:24FE}) functional $\Phi$ turns into a generating one of quantum averages of time-normally ordered products (time-normal averages, for short) of the {Heisenberg}\ operators conditional on the sources\/}. This unifies the kinematical analyses of \mbox{Refs.\ \protect\cite{APII,APIII}} and the dynamical ones of \mbox{Refs.\ \protect\cite{WickCaus,QDynResp}}. 

Setting $\eta (x,t)=\mu (x,t)=0$ in (\ref{eq:54XX}) we have, 
%=============================================
{\begin{multline}\hspace{0.4\columnwidth}\hspace{-0.4\twocolumnwidth} 
\Phi\protect\ensuremath{\big(
0 ,0,0, \zeta ,\nu ,\nu ^*
\big | 
J_{\mathrm{e}},
D_{\mathrm{e}},D_{\mathrm{e}}^*,
A_{\mathrm{e}},
E_{\mathrm{e}},E_{\mathrm{e}}^* 
 \big)} \\ 
= \Phi_{\mathrm{dev}}\protect\ensuremath{\big(
\zeta ,\nu ,\nu^* 
\big | 
A_{\mathrm{ext}},E_{\mathrm{ext}},E_{\mathrm{ext}}^*
 \big)} . 
\hspace{0.4\columnwidth}\hspace{-0.4\twocolumnwidth} 
\label{eq:25FF} % \nonumber % \Z 
\end{multline}}%
%+++++++++++++++++++++++++++++++++++++++++++++
This relation shows that, unlike fields, currents and dipoles depend only on the natural combinations (\ref{eq:26FH}). Comparing it to eq.\ (\protect\ref{eq:61HV}) we find, 
%=============================================
{\begin{multline}\hspace{0.4\columnwidth}\hspace{-0.4\twocolumnwidth} 
\Phi_{\mathrm{dev}}\protect\ensuremath{\big(
\zeta ,\nu ,\nu^* 
\big | 
A_{\mathrm{ext}},E_{\mathrm{ext}},E_{\mathrm{ext}}^*
 \big)} \\ 
= 
\protect\protect\ensuremath{\Big\langle {\mathcal T}{\mbox{\rm\boldmath$:$}}\exp\protect\ensuremath{\big(
i\zeta\protect{\hat{\mathcal J}}+i\nu^* \protect{\hat{\mathcal D}}-i\nu\protect{\hat{\mathcal D}}^{\dag}
 \big)} 
{\mbox{\rm\boldmath$:$}}
\Big\rangle} . 
\hspace{0.4\columnwidth}\hspace{-0.4\twocolumnwidth} 
\label{eq:28FK} % \nonumber % \Z 
\end{multline}}%
%+++++++++++++++++++++++++++++++++++++++++++++
In turn, this allows us to rewrite eq.\ (\protect\ref{eq:54XX}) as a relation between time-normal averages of the field, dipole and current operators, 
%=============================================
{\begin{multline} 
\protect\protect\ensuremath{\Big\langle {\mathcal T}{\mbox{\rm\boldmath$:$}}\exp\protect\ensuremath{\big(
i\eta\protect{\hat{\mathcal A}}+i\mu^*\protect{\hat{\mathcal E}}-i\mu\protect{\hat{\mathcal E}}^{\dag}
+i\zeta\protect{\hat{\mathcal J}}+i\nu^* \protect{\hat{\mathcal D}}-i\nu\protect{\hat{\mathcal D}}^{\dag}
 \big)} 
{\mbox{\rm\boldmath$:$}}
\Big\rangle} \\ 
= 
\protect\protect\ensuremath{\Big\langle {\mathcal T}{\mbox{\rm\boldmath$:$}}\exp \protect\protect\ensuremath{\big[
i\eta\Delta_{\text{R}}\protect\ensuremath{\big(
\protect{\hat{\mathcal J}}+J_{\textrm{e}}
 \big)} +i\mu^*G_{\text{R}}\protect\ensuremath{\big(
\protect{\hat{\mathcal D}}+D_{\mathrm{e}}
 \big)} \\ 
-i\mu G_{\text{R}}^*\protect\ensuremath{\big(
\protect{\hat{\mathcal D}}^{\dag}+D_{\mathrm{e}}^*
 \big)}
+i\zeta\protect{\hat{\mathcal J}}+i\nu^* \protect{\hat{\mathcal D}}-i\nu\protect{\hat{\mathcal D}}^{\dag}
\big]} 
{\mbox{\rm\boldmath$:$}}
\Big\rangle} . 
\label{eq:40FX} % \nonumber % \Z 
\end{multline}}%
%+++++++++++++++++++++++++++++++++++++++++++++
We moved the c-number factors inside the time-normal average. 
% hide 
Formula (\ref{eq:40FX}) is the starting point of all analyses in this paper. 

%********************************************************
\subsection{Classical phenomenology of radiation scattering}%
\label{ch:PX}
%*******************************************
\begin{figure}[t]
\begin{center}
\includegraphics[width=190pt]{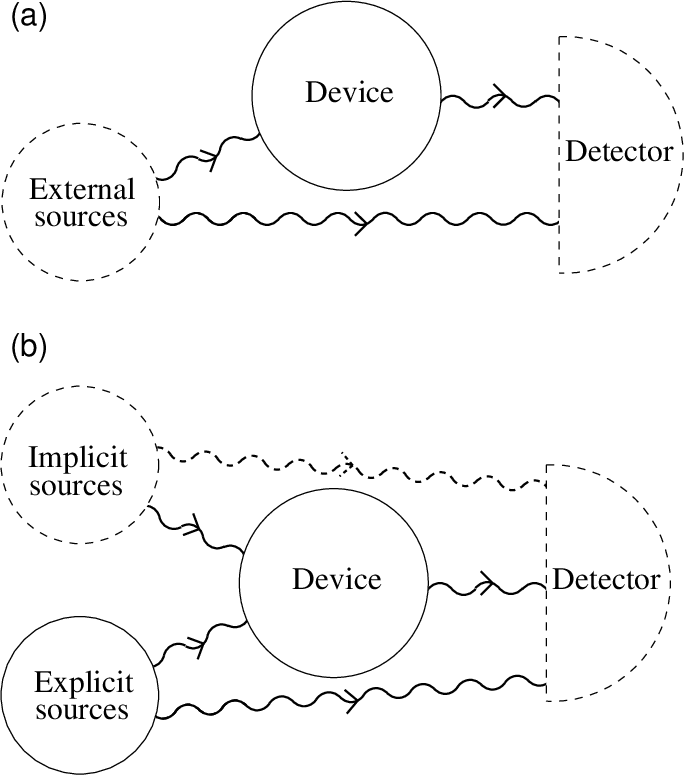}
\end{center}
\caption{Schematics of eqs.\ (\protect\ref{eq:2EF}) and (\ref{eq:41FY}). (a) In eq.\ (\protect\ref{eq:2EF}), radiation of external sources is scattered by a device, and full radiation is measured by a detector. (b) In eq.\ (\protect\ref{eq:41FY}), the sources are initially divided into implicit and explicit ones, and the detector is insensitive to the radiation of the implicit sources. To achieve full correspondence between the classical and quantum formulations, radiation of the implicit sources is added to the detected field by hand. Radiation of implicit sources which does ot does not reach the detector is shown schematically as a dashed wavy line, while all other types of radiation --- as solid wavy lines. Devices drawn with dashed lines occur implicitly.}
\label{fig:Scatt}
\end{figure}
%*******************************************
As a yardstick for quantum interactions, consider a classical scattering problem depicted in Fig.\ \ref{fig:Scatt}a. For simplicity we confine our discussion to the nonresonant\ field and current (the broad-band\ case). The general case will be restored in section \ref{ch:PE}. In the arrangement in Fig.\ \ref{fig:Scatt}a, radiation of some external sources $A_{\mathrm{ext}}(x,t)$ is incident on a device. Full radiation $A_{\mathrm{tot}}(x,t)$ seen by a detector includes $A_{\mathrm{ext}}(x,t)$ and radiation of the device, 
%=============================================
\protect{\begin{align}{{
 \begin{aligned} 
A_{\mathrm{tot}}(x,t) = A_{\mathrm{ext}}(x,t)+\int dt' \Delta_{\text{R}}(x,x',t-t')J(x',t'). 
\end{aligned}}}%
\label{eq:96EA} % \nonumber % \Z 
\end{align}}%
%+++++++++++++++++++++++++++++++++++++++++++++
The random current $J(x,t)$ describes the device. Sources of $A_{\mathrm{ext}}(x,t)$ and the detector occur implicitly; in Fig.\ \ref{fig:Scatt}a, they are drawn with dashed lines. 

External radiation is by definition regular (nonstochastic). The only source of randomness is stochasticity of the current $J(x,t)$. Its most general characterisation is given by a {\em conditional functional probability distribution\/}, 
%=============================================
\protect{\begin{align}{{
 \begin{aligned} 
p\protect\ensuremath{\big(
 J\big | A_{\mathrm{ext}}
 \big)} \geq 0. 
\end{aligned}}}%
\label{eq:63YH} % \nonumber % \Z 
\end{align}}%
%+++++++++++++++++++++++++++++++++++++++++++++
We stress that $p\protect\ensuremath{\big(
 J\big | A_{\mathrm{ext}}
 \big)}$ is a functional of two c-number functions $J(x,t)$ and $A_{\mathrm{ext}}(x,t)$ and not a function of two scalar variables. 
An alternative characterisation of the device is given by the generating functional of stochastic averages of the random current conditional on the incident field, 
%=============================================
\protect{\begin{align} 
& \overline{\hspace{0.1ex}J(x_1,t_1)\cdots J(x_m,t_m)\hspace{0.1ex}}\settoheight{\auxlv}{$|$}%
\raisebox{-0.3\auxlv}{$|_{A_{\mathrm{ext}}}$} 
= \frac{\delta ^m \Phi_{\mathrm{dev}}\protect\ensuremath{\big(
\zeta \big | A_{\mathrm{ext}}
 \big)}}{\delta \zeta (x_1,t_1)\cdots\delta \zeta (x_m,t_m)}, 
\nonumber % \Z 
\\ 
& \Phi_{\mathrm{dev}}\protect\ensuremath{\big(
\zeta \big | A_{\mathrm{ext}}
 \big)} = \overline{\hspace{0.1ex}
\exp \protect\protect\ensuremath{\bigg[
i \int dt \zeta (x,t) J(x,t)
\bigg]}\hspace{0.1ex}}\settoheight{\auxlv}{$\Big|$}%
\raisebox{-0.3\auxlv}{$\Big|_{A_{\mathrm{ext}}}$} . 
\label{eq:99ED} 
\end{align}}%
%+++++++++++++++++++++++++++++++++++++++++++++
For a general non-Markovian system, the conditional average $\overline{\hspace{0.1ex}\cdots\hspace{0.1ex}}\settoheight{\auxlv}{$|$}%
\raisebox{-0.3\auxlv}{$|_{A_{\mathrm{ext}}}$}$
is written explicitly as a path integral, 
%=============================================
\protect{\begin{align}{{
 \begin{aligned} 
\Phi_{\mathrm{dev}}\protect\ensuremath{\big(
\zeta \big | A_{\mathrm{ext}}
 \big)} 
= \prod_{x,t}\bigg\{\int d J(x,t)\bigg\}p\protect\ensuremath{\big(
 J\big | A_{\mathrm{ext}}
 \big)}\exp\protect\ensuremath{\big(
i\zeta J
 \big)} . 
\end{aligned}}}%
\label{eq:1EE} % \nonumber % \Z 
\end{align}}%
%+++++++++++++++++++++++++++++++++++++++++++++
We again resort to condensed notation (\ref{eq:3VS}). 

We do not introduce path integrals formally, thinking of them as multidimensional integrals in discretised time. 
This makes their algebraic manipulation straightforward. 
In particular, inverting the multidimensional Fourier-transformation (\ref{eq:1EE}) 
we find the formula, 
%=============================================
{\begin{multline}\hspace{0.4\columnwidth}\hspace{-0.4\twocolumnwidth} 
p\protect\ensuremath{\big(
 J\big | A_{\mathrm{ext}}
 \big)} = \prod_{x,t}\bigg\{\frac{dtdx}{2\pi }\int d \zeta (x,t)\bigg\} 
\exp\protect\ensuremath{\big(
- i \zeta J
 \big)} 
\\ \times 
\Phi_{\mathrm{dev}}\protect\ensuremath{\big(
\zeta \big | A_{\mathrm{ext}}
 \big)} , 
\hspace{0.4\columnwidth}\hspace{-0.4\twocolumnwidth} 
\label{eq:98EC} % \nonumber % \Z 
\end{multline}}%
%+++++++++++++++++++++++++++++++++++++++++++++
cf.\ eq.\ (\protect\ref{eq:55DT}) in the appendix. The infinitesimal scaling factor $\Pi_{x,t}dxdt$ emphasises that our formulae are only symbolic. For more details see section \ref{ch:R} and the appendix.

Full characterisation of the scattering experiment in Fig.\ \ref{fig:Scatt}a is given by the generating functional of joint stochatic averages of the field and current, 
%=============================================
{\begin{multline}\hspace{0.4\columnwidth}\hspace{-0.4\twocolumnwidth} 
\overline{\hspace{0.1ex} {
\exp\protect\ensuremath{\big(i\eta A_{\mathrm{tot}}
+i\zeta J
 \big)}
} 
\hspace{0.1ex}}\settoheight{\auxlv}{$|$}%
\raisebox{-0.3\auxlv}{$|_{A_{\mathrm{ext}}}$} 
\\ 
= \overline{\hspace{0.1ex}\exp 
\protect\protect\ensuremath{\big[
i\eta\protect\ensuremath{\big(
A_{\mathrm{ext}}+\Delta_{\text{R}} J
 \big)} 
+ i\zeta J 
\big]} 
\hspace{0.1ex}}\settoheight{\auxlv}{$|$}%
\raisebox{-0.3\auxlv}{$|_{A_{\mathrm{ext}}}$} . 
\hspace{0.4\columnwidth}\hspace{-0.4\twocolumnwidth} 
\label{eq:15EV} % \nonumber % \Z 
\end{multline}}%
%+++++++++++++++++++++++++++++++++++++++++++++
The RHS here may also be written explicitly as a functional integral, 
%=============================================
{\begin{multline}\hspace{0.4\columnwidth}\hspace{-0.4\twocolumnwidth} 
\overline{\hspace{0.1ex} {
\exp\protect\ensuremath{\big(i\eta A_{\mathrm{tot}}
+i\zeta J
 \big)}
} 
\hspace{0.1ex}}\settoheight{\auxlv}{$|$}%
\raisebox{-0.3\auxlv}{$|_{A_{\mathrm{ext}}}$} 
= \prod_{x,t}\bigg\{\int d J(x,t)\bigg\}
p\protect\ensuremath{\big(
 J\big | A_{\mathrm{ext}}
 \big)}
\\ \times 
\exp 
\protect\protect\ensuremath{\big[
i\eta\protect\ensuremath{\big(
A_{\mathrm{ext}}+\Delta_{\text{R}} J
 \big)} 
+ i\zeta J 
\big]} . 
\hspace{0.4\columnwidth}\hspace{-0.4\twocolumnwidth} 
\label{eq:2EF} % \nonumber % \Z 
\end{multline}}%
%+++++++++++++++++++++++++++++++++++++++++++++
These formulae express two simple facts: that the full field is external field plus radiation of the device, cf.\ eq.\ (\protect\ref{eq:96EA}), and that the sole source of stochasticity is randomness of $J(x,t)$. 
%********************************************************
\subsection{Quantum {\em versus\/} classical field-scattering problem }%
\label{ch:GS}
%********************************************************
It is instructive to compare the classical eq.\ (\protect\ref{eq:15EV}) to the quantum formula (\ref{eq:40FX}). Confining the latter for simplicity to the broad-band\ field and current we obtain, 
%=============================================
{\begin{multline}\hspace{0.4\columnwidth}\hspace{-0.4\twocolumnwidth} 
\protect\protect\ensuremath{\Big\langle {\mathcal T}{\mbox{\rm\boldmath$:$}}\exp\protect\ensuremath{\big(
i\eta\protect{\hat{\mathcal A}}+i\zeta\protect{\hat{\mathcal J}}
 \big)} 
{\mbox{\rm\boldmath$:$}}
\Big\rangle} \\ 
= 
\protect\protect\ensuremath{\Big\langle {\mathcal T}{\mbox{\rm\boldmath$:$}}\exp \protect\protect\ensuremath{\big[
i\eta 
\Delta_{\text{R}}\protect\ensuremath{\big(
\protect{\hat{\mathcal J}}+J_{\mathrm{e}}
 \big)}+i\zeta\protect{\hat{\mathcal J}} 
\big]} 
{\mbox{\rm\boldmath$:$}}
\Big\rangle} . 
\hspace{0.4\columnwidth}\hspace{-0.4\twocolumnwidth} 
\label{eq:78JP} % \nonumber % \Z 
\end{multline}}%
%+++++++++++++++++++++++++++++++++++++++++++++
Reduced to the current modes, eq.\ (\protect\ref{eq:28FK}) reads, 
%=============================================
\protect{\begin{align}{{
 \begin{aligned} 
\Phi_{\mathrm{dev}}\protect\ensuremath{\big(
\zeta 
\big | 
A_{\mathrm{ext}} 
 \big)} %\\ 
= 
\protect\protect\ensuremath{\Big\langle {\mathcal T}{\mbox{\rm\boldmath$:$}}\exp\protect\ensuremath{\big(
i\zeta\protect{\hat{\mathcal J}}
 \big)} 
{\mbox{\rm\boldmath$:$}}
\Big\rangle} . 
\end{aligned}}}%
\label{eq:53HM} % \nonumber % \Z 
\end{align}}%
%+++++++++++++++++++++++++++++++++++++++++++++
We see that in quantum mechanics things are a trifle more complicated than in classical mechanics. While properties of the current operator depend on the full external field $A_{\mathrm{ext}}(x,t)$, those of the field operator depend on $A_{\mathrm{ext}}(x,t)$ through the current operator and separately on $J_{\mathrm{e}}(x,t)$ through the factor 
%=============================================
\protect{\begin{align}{{
 \begin{aligned} 
\exp\protect\ensuremath{\big(
i\eta \Delta_{\text{R}} J_{\mathrm{e}}
 \big)} 
\end{aligned}}}%
\label{eq:79JQ} % \nonumber % \Z 
\end{align}}%
%+++++++++++++++++++++++++++++++++++++++++++++
in eq.\ (\protect\ref{eq:78JP}). We therefore multiply eq.\ (\protect\ref{eq:78JP}) by the additional factor, 
%=============================================
\protect{\begin{align}{{
 \begin{aligned} 
\exp\protect\ensuremath{\big(
i\eta A_{\textrm{e}}
 \big)} . 
\end{aligned}}}%
\label{eq:80JR} % \nonumber % \Z 
\end{align}}%
%+++++++++++++++++++++++++++++++++++++++++++++
The resulting quantum formula reads, 
%=============================================
{\begin{multline}\hspace{0.4\columnwidth}\hspace{-0.4\twocolumnwidth} 
\protect\protect\ensuremath{\Big\langle {\mathcal T}{\mbox{\rm\boldmath$:$}}\exp \protect\protect\ensuremath{\big[
i\eta\protect\ensuremath{\big(
\protect{\hat{\mathcal A}}+A_{\textrm{e}}
 \big)} +i\zeta\protect{\hat{\mathcal J}}
\big]} 
{\mbox{\rm\boldmath$:$}}
\Big\rangle} \\ 
= 
\protect\protect\ensuremath{\Big\langle {\mathcal T}{\mbox{\rm\boldmath$:$}}\exp \protect\protect\ensuremath{\big[
i\eta\protect\ensuremath{\big(
A_{\textrm{ext}}+\Delta_{\text{R}}\protect{\hat{\mathcal J}}
 \big)}+i\zeta\protect{\hat{\mathcal J}} 
\big]} 
{\mbox{\rm\boldmath$:$}}
\Big\rangle} , 
\hspace{0.4\columnwidth}\hspace{-0.4\twocolumnwidth} 
\label{eq:41FY} % \nonumber % \Z 
\end{multline}}%
%+++++++++++++++++++++++++++++++++++++++++++++
where use was made of the obvious relation, 
%=============================================
\protect{\begin{align}{{
 \begin{aligned} 
\exp\protect\ensuremath{\big(
i\eta \Delta_{\text{R}} J_{\mathrm{e}}
 \big)}\exp\protect\ensuremath{\big(
i\eta A_{\textrm{e}}
 \big)} = \exp\protect\ensuremath{\big(
i\eta A_{\textrm{ext}}
 \big)} . 
\end{aligned}}}%
\label{eq:81JS} % \nonumber % \Z 
\end{align}}%
%+++++++++++++++++++++++++++++++++++++++++++++
In eq.\ (\protect\ref{eq:41FY}), the RHS and hence the LHS depend only on the full external field $A_{\mathrm{ext}}(x,t)$. 

Comparing eq.\ (\protect\ref{eq:15EV}) and (\ref{eq:41FY}) we see that they coincide up to the replacement of operators by c-numbers, 
and of the time-normal averages by classical stochastic averages, 
%=============================================
\protect{\begin{align}{{
 \begin{aligned} 
& \protect{\hat{\mathcal A}}(x,t)+A_{\textrm{e}}(x,t)\leftrightarrow A_{\mathrm{tot}}(x,t), & 
& \protect{\hat{\mathcal J}}(x,t)\leftrightarrow J(x,t), \\ 
& \protect\protect\ensuremath{\langle 
{\mathcal T}{\mbox{\rm\boldmath$:$}}(\cdots){\mbox{\rm\boldmath$:$}}
\rangle} \leftrightarrow \overline{\hspace{0.1ex}(\cdots)\hspace{0.1ex}}. 
\end{aligned}}}%
\label{eq:95DZ} % \nonumber % \Z 
\end{align}}%
%+++++++++++++++++++++++++++++++++++++++++++++
The field operator $\protect{\hat{\mathcal A}}(x,t)$ thus corresponds not to the full field $A_{\mathrm{tot}}(x,t)$, but to the {\em radiated field\/} $A(x,t)$, 
%=============================================
{\begin{multline}\hspace{0.4\columnwidth}\hspace{-0.4\twocolumnwidth} 
 A(x,t) = 
 A_{\mathrm{tot}}(x,t) - A_{\mathrm{e}}(x,t) 
\\ 
= \int dt' \Delta_{\text{R}}(x,x',t-t')\protect\protect\ensuremath{\big[
J(x',t') + J_{\mathrm{e}}(x',t') 
\big]} . 
\hspace{0.4\columnwidth}\hspace{-0.4\twocolumnwidth} 
\label{eq:65YK} % \nonumber % \Z 
\end{multline}}%
%+++++++++++++++++++++++++++++++++++++++++++++
The latter is measured in the experimental arrangement shown in Fig.\ \ref{fig:Scatt}b. All external sources are divided into implicit and explicit ones. Implicit sources are responsible for the field $A_{\mathrm{e}}(x,t)$. This field affects the device but not the detector. Explicit sources are described by the current $J_{\mathrm{e}}(x,t)$. Radiation of the latter affects both the device and the detector. 

That the detector sees radiation of some sources and does not see radiation of others may seem unnatural, but it reflects the situation in quantum mechanics where quantised fields and c-number sources are objects of different nature. In fact, whether the detector does or does not see $A_{\mathrm{e}}(x,t)$ is an additional assumption to be made in a detection model. We return to this question elsewhere. 
%********************************************************
\subsection{Cancellation of the in-field}%
\label{ch:PI}
%********************************************************
The message of eq.\ (\protect\ref{eq:78JP}) is that, {\em under the time-normal averaging, classical radiation laws apply directly to {Heisenberg}\ operators\/}. That is, in a time-normal average (and {\em only\/} in a time-normal average) we can write, 
%=============================================
{\begin{multline}\hspace{0.4\columnwidth}\hspace{-0.4\twocolumnwidth} 
\protect{\hat{\mathcal A}}(x,t) = \int dx' dt' \Delta_{\text{R}}(x,x',t-t') 
\\ \times 
\protect\protect\ensuremath{\big[
\protect{\hat{\mathcal J}}(x',t')+J_{\textrm{e}}(x',t')
\big]} . 
\hspace{0.4\columnwidth}\hspace{-0.4\twocolumnwidth} 
\label{eq:35FS} % \nonumber % \Z 
\end{multline}}%
%+++++++++++++++++++++++++++++++++++++++++++++
It is instructive to compare this relation to the standard quantum-field-theoretical formula connecting the {Heisenberg}\ and free-field operators. 
Without the source, (\ref{eq:35FS}) becomes, 
%=============================================
\protect{\begin{align}{{
 \begin{aligned} 
\protect{\hat{\mathcal A}}(x,t) & = \int dx' dt' \Delta_{\text{R}}(x,x',t-t') \protect{\hat{\mathcal J}}(x',t') . 
\end{aligned}}}%
\label{eq:37FU} % \nonumber % \Z 
\end{align}}%
%+++++++++++++++++++++++++++++++++++++++++++++
As a Hilbert-space formula, this relation cannot be correct because it does not preserve commutational relations. The right formula \cite{Schweber} should include the free-field operator (in-field), 
%=============================================
{\begin{multline}\hspace{0.4\columnwidth}\hspace{-0.4\twocolumnwidth} 
\protect{\hat{\mathcal A}}(x,t) = \hat A(x,t) 
\\ \ \ \ + \int dx' dt' \Delta_{\text{R}}(x,x',t-t') \protect{\hat{\mathcal J}}(x',t') . 
\hspace{0.4\columnwidth}\hspace{-0.4\twocolumnwidth} 
\label{eq:73JJ} % \nonumber % \Z 
\end{multline}}%
%+++++++++++++++++++++++++++++++++++++++++++++
Under the time-normal averaging, the in-field cancels. This is partly due to the vacuum initial state of the field, but only partly. 
The in-field operator $\hat A(x,t)$ does not commute with $\protect{\hat{\mathcal A}}(x,t)$ and $\protect{\hat{\mathcal J}}(x,t)$, so that its disappearance under the time-normal averaging in eq.\ (\protect\ref{eq:78JP}) is anything but trivial. 

%********************************************************
\section{Conditional P-functional}%
\label{ch:P}
%********************************************************
\subsection{Conditional time-normal quasiprobability distribution of the quantum current}%
\label{ch:PQ}
%********************************************************
In conventional phase-space approaches \cite{MandelWolf,AgarwalWolf}, each type of operator ordering is associated with a corresponding type of quasidistribution. Formally, quasidistributions may be defined postulating that the relation between quantum averages of operators ordered in a particular way and the associated quasidistribution emulates the classical relation between stochastic averages and probability distributions. Applying this idea to interacting systems, it is natural to introduce {\em conditional functional time-normal quasiprobability distributions\/}, or {\em conditional P-functionals\/}, of quantum dynamical variables. By definition, they are related to time-normal averages of these variables by formulae emulating classical relations between multitime stochastic averages and corresponding functional probability distributions. Conditional P-functionals thus generalise two concepts: that of conditional functional probability distribution to quantum mechanics, and that of P-function to Heisenberg fields. 

So, postulating eq.\ (\protect\ref{eq:98EC}) for the quantum $\Phi_{\mathrm{dev}}$ given by (\ref{eq:53HM}), we define the conditional P-functional of the quantum current as, 
%=============================================
{\begin{multline}\hspace{0.4\columnwidth}\hspace{-0.4\twocolumnwidth} 
p\protect\ensuremath{\big(
 J\big | A_{\mathrm{ext}}
 \big)} = \prod_{t}\bigg\{\frac{dtdx}{2\pi }\int d \zeta (t)\bigg\} 
\\ \times 
\protect\protect\ensuremath{\bigg\langle 
{\mathcal T}{\mbox{\rm\boldmath$:$}}\exp \protect\protect\ensuremath{\bigg\{
i \int dxdt \zeta (x,t) \protect\protect\ensuremath{\big[
\protect{\hat{\mathcal J}}(x,t) - J(x,t)
\big]} 
\bigg\}} 
{\mbox{\rm\boldmath$:$}}
\bigg\rangle}%\cnd{\Big}{A_{\mathrm{e}}} 
. 
\hspace{0.4\columnwidth}\hspace{-0.4\twocolumnwidth} 
\label{eq:83ZD} % \nonumber % \Z 
\end{multline}}%
%+++++++++++++++++++++++++++++++++++++++++++++
The inverse relation emulates eq.\ (\protect\ref{eq:1EE}): 
\begin{widetext} 
%=============================================
\protect{\begin{align}{{
 \begin{aligned} 
\protect\protect\ensuremath{\bigg\langle 
{\mathcal T}{\mbox{\rm\boldmath$:$}}\exp \protect\protect\ensuremath{\bigg[
i \int dt \zeta (x,t)\protect{\hat{\mathcal J}}(x,t)\bigg]} 
{\mbox{\rm\boldmath$:$}}
\bigg\rangle} %\\ 
= \prod_{x,t}\bigg\{\int d J(x,t)\bigg\}p\protect\ensuremath{\big(
 J\big | A_{\mathrm{ext}}
 \big)} 
\exp \protect\protect\ensuremath{\bigg[
i \int dxdt \zeta (x,t) J(x,t)
\bigg]} . 
\end{aligned}}}%
\label{eq:19EZ} % \nonumber % \Z 
\end{align}}%
%+++++++++++++++++++++++++++++++++++++++++++++
Note that the logic here is the other way around compared to eqs.\ (\protect\ref{eq:1EE}), (\ref{eq:98EC}). The primary quantity is functional (\ref{eq:53HM}), $p\protect\ensuremath{\big(
 J\big | A_{\mathrm{ext}}
 \big)}$ is defined by eq.\ (\protect\ref{eq:83ZD}), while eq.\ (\protect\ref{eq:19EZ}) is found inverting the latter. 

Using eq.\ (\protect\ref{eq:19EZ}), eq.\ (\protect\ref{eq:40FX}) may be written in the form, 
%\begin{widetext} 
%=============================================
{\begin{multline} 
\protect\protect\ensuremath{\bigg\langle {\mathcal T}{\mbox{\rm\boldmath$:$}}
\exp\protect\ensuremath{\bigg(i\int dxdt 
\protect\protect\ensuremath{\Big\{
\eta(x,t)\protect\protect\ensuremath{\big[
\protect{\hat{\mathcal A}}(x,t)+A_{\mathrm{e}}(x,t)
\big]} %\\
+\zeta(x,t)\protect{\hat{\mathcal J}}(x,t)
\Big\}} 
 \bigg)}
{\mbox{\rm\boldmath$:$}} 
\bigg\rangle}% \cnd{\Big}{A_{\mathrm{e}},J_{\mathrm{e}}} 
\\ 
= \prod_{x,t}\bigg\{\int d J(x,t)\bigg\}
p\protect\ensuremath{\big(
 J\big | A_{\mathrm{ext}}
 \big)}
%\\ \times 
\exp\protect\ensuremath{\bigg(i\int dxdt 
\protect\protect\ensuremath{\bigg\{
\zeta(x,t)J(x,t)
\\ 
+ \eta(x,t) \protect\protect\ensuremath{\bigg[
A_{\mathrm{ext}}(x,t) 
+\int dx'dt' \Delta_{\text{R}}(t-t')J(x',t')
\bigg]} 
\bigg\}} 
 \bigg)} . 
\label{eq:5EK} % \nonumber % \Z 
\end{multline}}%
%+++++++++++++++++++++++++++++++++++++++++++++
%\end{widetext}%
This is a quantum analog of eq.\ (\protect\ref{eq:2EF}). It differs from the latter in replacements (\ref{eq:95DZ}), and in that the P-functional $p\protect\ensuremath{\big(
 J\big | A_{\mathrm{ext}}
 \big)}$ needs not be nonnegative. 

Reality and causality properties of the P-functionals are inherited from those of the the time-normal averages (cf.\ section \ref{ch:TNC}). Using reality of the latter it is straightforward to show that the P-functionals are also real. We avoid formulating causality conditions for the P-functionals which are not transparent. It suffices to say that {\em causality properties of the P-functionals coincide with those of the functional probability distributions\/}. 

Equation (\ref{eq:5EK}) may be extended to a full quantum treatment replacing, 
%=============================================
\protect{\begin{align}{{
 \begin{aligned} 
A_{\mathrm{e}}(x,t)\to A_{\mathrm{e}}(x,t)+ a_{\mathrm{e}}(x,t). 
\end{aligned}}}%
\label{eq:72JH} % \nonumber % \Z 
\end{align}}%
%+++++++++++++++++++++++++++++++++++++++++++++
In classical mechanics, $ a_{\mathrm{e}}(x,t)$ does not exist. It appears only in quantum mechanics, where it reflects noncommutativity of the operators. $a_{\mathrm{e}}(x,t)$ and $A_{\mathrm{e}}(x,t)$ not just differ, but are of different nature: one is an auxiliary variable, and the other is an external source. It is a nontrivial property of quantum dynamics that they occur in the functional $\Phi $ as a sum. All quantum-classical correspondences we discuss in this paper are subject to two facts: absence of Planck's constant in dynamical relations in causal variables, and the consistency relation (\ref{eq:29FL}). In no way should importance of the latter be overlooked. 
%********************************************************
\subsection{Extension to the general case}%
\label{ch:PE}
%********************************************************
The main advantage of conditional P-functionals is that they allow for {\em doing quantum electrodynamics by thinking classically\/}. As an example, let us ``derive'' eq.\ (\protect\ref{eq:40FX}) from classical considerations and correspondence rules (\ref{eq:95DZ}). In the general case of Hamiltonian (\ref{eq:86FB}), eqs.\ (\protect\ref{eq:95DZ}) should be supplemented by correspondences for dipoles and optical fields, 
%=============================================
\protect{\begin{align}{{
 \begin{aligned} 
\protect{\hat{\mathcal E}}(x,t)& \leftrightarrow E(x,t), & 
\protect{\hat{\mathcal D}}(x,t)& \leftrightarrow D(x,t) , & 
\protect{\hat{\mathcal E}}^{\dag}(x,t)& \leftrightarrow E^*(x,t), & 
\protect{\hat{\mathcal D}}^{\dag}(x,t)& \leftrightarrow D^*(x,t) , 
\end{aligned}}}%
\label{eq:44HB} % \nonumber % \Z 
\end{align}}%
%+++++++++++++++++++++++++++++++++++++++++++++
where $E(x,t)$ is the radiated optical field, 
%=============================================
\protect{\begin{align}{{
 \begin{aligned} 
E(x,t) = \int dx'dt' G_{\text{R}}(x,x',t-t')\protect\protect\ensuremath{\big[
D(x,t) + D_{\textrm{e}}(x,t)
\big]} . 
\end{aligned}}}%
\label{eq:45HC} % \nonumber % \Z 
\end{align}}%
%+++++++++++++++++++++++++++++++++++++++++++++
The device is now formally described by two random quantities, current $J(x,t)$ and dipole $D(x,t)$. Their joint probablity distribution is conditional on the external fields (\ref{eq:26FH}), 
%=============================================
\protect{\begin{align}{{
 \begin{aligned} 
p\protect\ensuremath{\big(
J,D,D^*\big|A_{\text{ext}},E_{\text{ext}},E_{\text{ext}}^*
 \big)} \geq 0. 
\end{aligned}}}%
\label{eq:46HD} % \nonumber % \Z 
\end{align}}%
%+++++++++++++++++++++++++++++++++++++++++++++
Using it we can construct the characteristic functional of joint statistical averages of the currents and dipoles, 
%=============================================
{\begin{multline} 
\overline{\hspace{0.1ex}\exp\protect\ensuremath{\big(
i\zeta J +i\nu ^* D -i\nu D^*
 \big)} \hspace{0.1ex}} 
= \prod_{x,t}\bigg\{\int dJ(x,t)d^2D(x,t)\bigg\} 
\\ \times 
p\protect\ensuremath{\big(
J,D,D^*\big|A_{\text{ext}},E_{\text{ext}},E_{\text{ext}}^*
 \big)}\exp\protect\ensuremath{\big(
i\zeta J +i\nu ^* D -i\nu D^*
 \big)} . 
\label{eq:47HE} % \nonumber % \Z 
\end{multline}}%
%+++++++++++++++++++++++++++++++++++++++++++++
Using eqs.\ (\protect\ref{eq:65YK}) and (\ref{eq:45HC}), we also obtain the characteristic functional of joint statistical averages of the fields, currents and dipoles, 
%=============================================
{\begin{multline} 
\overline{\hspace{0.1ex}\exp\protect\ensuremath{\big(
i\eta A+i\mu ^*E -i\mu E^*
+i\zeta J +i\nu ^* D -i\nu D^*
 \big)} \hspace{0.1ex}} 
\\ 
= \prod_{x,t}\bigg\{\int dJ(x,t)d^2D(x,t)\bigg\}
p\protect\ensuremath{\big(
J,D,D^*\big|A_{\text{ext}},E_{\text{ext}},E_{\text{ext}}^*
 \big)} 
\\ \times 
\exp\protect\protect\ensuremath{\big[
i\eta \Delta_{\text{R}}\protect\ensuremath{\big(
J+J_{\textrm{e}}
 \big)}+i\mu ^*G_{\text{R}}\protect\ensuremath{\big(
D+D_{\textrm{e}}
 \big)} -i\mu G_{\text{R}}^*\protect\ensuremath{\big(
D^*+D_{\textrm{e}}^*
 \big)}
+i\zeta J +i\nu ^* D -i\nu D^*
\big]} . 
\label{eq:48HF} % \nonumber % \Z 
\end{multline}}%
%+++++++++++++++++++++++++++++++++++++++++++++
Comparing these two relations, we find the formula, 
%=============================================
{\begin{multline} 
\overline{\hspace{0.1ex}\exp\protect\ensuremath{\big(
i\eta A+i\mu ^*E -i\mu E^*
+i\zeta J +i\nu ^* D -i\nu D^*
 \big)} \hspace{0.1ex}} \\ 
= \overline{\hspace{0.1ex}\exp\protect\protect\ensuremath{\big[
i\eta \Delta_{\text{R}}\protect\ensuremath{\big(
J+J_{\textrm{e}}
 \big)}+i\mu ^*G_{\text{R}}\protect\ensuremath{\big(
D+D_{\textrm{e}}
 \big)} -i\mu G_{\text{R}}^*\protect\ensuremath{\big(
D^*+D_{\textrm{e}}^*
 \big)}
+i\zeta J +i\nu ^* D -i\nu D^*
\big]}\hspace{0.1ex}} . 
\label{eq:49HH} % \nonumber % \Z 
\end{multline}}%
%+++++++++++++++++++++++++++++++++++++++++++++
Applying replacements (\ref{eq:95DZ}), (\ref{eq:44HB}) to this relation we indeed recover eq.\ (\protect\ref{eq:40FX}). 

Consider now the logic of this ``derivation'' in more detail. Applying the said replacements to eq.\ (\protect\ref{eq:47HE}) is equivalent to defining the conditional P-functional, 
%=============================================
{\begin{multline} 
p\protect\ensuremath{\big(
J,D,D^*\big|A_{\text{ext}},E_{\text{ext}},E_{\text{ext}}^*
 \big)} = \prod_{x,t}\bigg\{\frac{(dtdx)^3}{2\pi ^3}\int d\zeta (x,t)d^2\nu (x,t)\bigg\} 
\\ \times 
\protect\protect\ensuremath{\Big\langle {\mathcal T}{\mbox{\rm\boldmath$:$}}\exp \protect\protect\ensuremath{\big[
i\zeta\protect\ensuremath{\big(
\protect{\hat{\mathcal J}}-J
 \big)} +i\nu^*\protect\ensuremath{\big(
\protect{\hat{\mathcal D}}-D
 \big)} -i\nu\protect\ensuremath{\big(
\protect{\hat{\mathcal D}}^{\dag}-D^*
 \big)} 
\big]} 
{\mbox{\rm\boldmath$:$}}
\Big\rangle} . 
\label{eq:51HK} % \nonumber % \Z 
\end{multline}}%
%+++++++++++++++++++++++++++++++++++++++++++++
Inverting this definition we find the quantum counterpart of (\ref{eq:47HE}), 
%=============================================
{\begin{multline} 
\protect\protect\ensuremath{\Big\langle {\mathcal T}{\mbox{\rm\boldmath$:$}}\exp\protect\ensuremath{\big(
i\zeta\protect{\hat{\mathcal J}}+i\nu^* \protect{\hat{\mathcal D}}-i\nu\protect{\hat{\mathcal D}}^{\dag}
 \big)} 
{\mbox{\rm\boldmath$:$}}
\Big\rangle} 
= \prod_{x,t}\bigg\{\int dJ(x,t)d^2D(x,t)\bigg\} 
\\ \times 
p\protect\ensuremath{\big(
J,D,D^*\big|A_{\text{ext}},E_{\text{ext}},E_{\text{ext}}^*
 \big)}\exp\protect\ensuremath{\big(
i\zeta J +i\nu ^* D -i\nu D^*
 \big)} . 
\label{eq:50HJ} % \nonumber % \Z 
\end{multline}}%
%+++++++++++++++++++++++++++++++++++++++++++++
Furthermore, applying the quantum-classical correspondences to eq.\ (\protect\ref{eq:48HF}) we obtain, 
%=============================================
{\begin{multline} 
\protect\protect\ensuremath{\Big\langle {\mathcal T}{\mbox{\rm\boldmath$:$}}\exp\protect\ensuremath{\big(
i\eta\protect{\hat{\mathcal A}}+i\mu^*\protect{\hat{\mathcal E}}-i\mu\protect{\hat{\mathcal E}}^{\dag}
+i\zeta\protect{\hat{\mathcal J}}+i\nu^* \protect{\hat{\mathcal D}}-i\nu\protect{\hat{\mathcal D}}^{\dag}
 \big)} 
{\mbox{\rm\boldmath$:$}}
\Big\rangle} \\ 
= \prod_{x,t}\bigg\{\int dJ(x,t)d^2D(x,t)\bigg\}
p\protect\ensuremath{\big(
J,D,D^*\big|A_{\text{ext}},E_{\text{ext}},E_{\text{ext}}^*
 \big)} 
\\ \times 
\exp\protect\protect\ensuremath{\big[
i\eta \Delta_{\text{R}}\protect\ensuremath{\big(
J+J_{\textrm{e}}
 \big)}+i\mu ^*G_{\text{R}}\protect\ensuremath{\big(
D+D_{\textrm{e}}
 \big)} -i\mu G_{\text{R}}^*\protect\ensuremath{\big(
D^*+D_{\textrm{e}}^*
 \big)}
+i\zeta J +i\nu ^* D -i\nu D^*
\big]} . 
\label{eq:52HL} % \nonumber % \Z 
\end{multline}}%
%+++++++++++++++++++++++++++++++++++++++++++++
\end{widetext}%
Comparing eqs.\ (\protect\ref{eq:50HJ}) and (\ref{eq:52HL}) we recover eq.\ (\protect\ref{eq:40FX}). However, there is no other way to actually {\em prove\/} (\ref{eq:52HL}) except by showing that it follows from (\ref{eq:40FX}), which in turn is another form of (\ref{eq:54XX}). Strictly speaking, eq.\ (\protect\ref{eq:52HL}) {\em is\/} eq.\ (\protect\ref{eq:54XX}) written using notation (\ref{eq:61HV}), (\ref{eq:28FK}) and definition (\ref{eq:51HK}). One may say that eq.\ (\protect\ref{eq:52HL}) is a {\em mnemonic form\/} of the quantum relation (\ref{eq:54XX}): classical connotations of the former allow one to easily memorise it. The actual {\em derivation\/} of eqs.\ (\protect\ref{eq:54XX}), (\ref{eq:40FX}) and (\ref{eq:52HL}) is that given in \mbox{Refs.\ \protect\cite{WickCaus,QDynResp}}. All we do here is rewrite the obscure eq.\ (\protect\ref{eq:54XX}) in a series of physically more transparent forms. 

%********************************************************
\section{Self-action problem in terms of P-functionals}%
\label{ch:PS}
%********************************************************
Conditional P-functionals also give a natural description of the electromagnetic self-action, or ``dressing,'' problem. Again, we start from a more compact broad-band\ case. Reduced to fields and currents, the dressing relation (\ref{eq:91FH}) becomes, 
 
%=============================================
\protect{\begin{align}{{
 \begin{aligned} 
\Phi_{\mathrm{dev}}\protect\ensuremath{\big(
\zeta \big| A_{\text{ext}}
 \big)} 
= 
\exp\protect\ensuremath{\bigg(
-i\frac{\delta }{\delta A_{\mathrm{ext}}}
\Delta_{\text{R}}
\frac{\delta }{\delta \zeta }
 \bigg)} 
\Phi_{\mathrm{dev}}^{\mathrm{I}}\protect\ensuremath{\big(
\zeta \big| A_{\text{ext}}
 \big)} , 
\end{aligned}}}%
\label{eq:23YA} % \nonumber % \Z 
\end{align}}%
%+++++++++++++++++++++++++++++++++++++++++++++
This formula implies the response definition of $\Phi_{\mathrm{dev}}^{\mathrm{I}}$ by eq.\ (\protect\ref{eq:64HY}). Following the pattern of eqs.\ (\protect\ref{eq:83ZD}), (\ref{eq:19EZ}) we define, 
%=============================================
{\begin{multline}\hspace{0.4\columnwidth}\hspace{-0.4\twocolumnwidth} 
p\protect\ensuremath{\big(
 J\big | A_{\mathrm{ext}}
 \big)} = \prod_{x,t}\bigg\{\frac{dxdt}{2\pi }\int d \zeta (x,t)\bigg\} 
\\ \times 
\text{Tr} \hat \rho _{\textrm{dev}}{ 
{\mathcal T}{\mbox{\rm\boldmath$:$}}\exp \protect\protect\ensuremath{\big[
i \zeta\protect\ensuremath{\big(
\hat J' - J
 \big)} 
\big]} 
{\mbox{\rm\boldmath$:$}}
} 
, 
\hspace{0.4\columnwidth}\hspace{-0.4\twocolumnwidth} 
\label{eq:56HQ} % \nonumber % \Z 
\end{multline}}%
%+++++++++++++++++++++++++++++++++++++++++++++
and 
%=============================================
{\begin{multline}\hspace{0.4\columnwidth}\hspace{-0.4\twocolumnwidth} 
\Phi_{\mathrm{dev}}^{\mathrm{I}}\protect\ensuremath{\big(
\zeta \big| A_{\text{ext}}
 \big)} = \text{Tr} \hat\rho _{\textrm{dev}}{\mathcal T}{\mbox{\rm\boldmath$:$}}\exp\protect\ensuremath{\big(
i\zeta \hat J'
 \big)} {\mbox{\rm\boldmath$:$}} \\ 
= \prod_{x,t}\bigg\{\int d J(x,t)\bigg\}p^{\mathrm{I}}\protect\ensuremath{\big(
 J\big | A_{\mathrm{ext}}
 \big)} 
\exp\protect\ensuremath{\big(
i \zeta J
 \big)} . 
\hspace{0.4\columnwidth}\hspace{-0.4\twocolumnwidth} 
\label{eq:89ZL} % \nonumber % \Z 
\end{multline}}%
%+++++++++++++++++++++++++++++++++++++++++++++
$\hat J'(x,t)$ was defined in section \ref{ch:UPS}. 
We use condensed notation (\ref{eq:3VS}). 
The meaning of $p^{\mathrm{I}}$ is yet to be understood. Substituting (\ref{eq:89ZL}) into the dressing formula (\ref{eq:23YA}) we have, 
%=============================================
{\begin{multline}\hspace{0.4\columnwidth}\hspace{-0.4\twocolumnwidth} 
\Phi_{\mathrm{dev}}\protect\ensuremath{\big(
\zeta \big| A_{\mathrm{ext}}
 \big)} = 
\prod_{x,t}\bigg\{\int d J(x,t)\bigg\}
\\ \times 
\exp\protect\ensuremath{\bigg(
-i\frac{\delta }{\delta A_{\mathrm{ext}}}
\Delta_{\text{R}}
\frac{\delta }{\delta \zeta }
 \bigg)} 
p^{\mathrm{I}}\protect\ensuremath{\big(
 J\big | A_{\mathrm{ext}}
 \big)} 
\exp\protect\ensuremath{\big(i \zeta J
 \big)} 
. 
\hspace{0.4\columnwidth}\hspace{-0.4\twocolumnwidth} 
\label{eq:84ZE} % \nonumber % \Z 
\end{multline}}%
%+++++++++++++++++++++++++++++++++++++++++++++
The integrand here is transformed in two steps: 
%=============================================
{\begin{multline}\hspace{0.4\columnwidth}\hspace{-0.4\twocolumnwidth} 
\exp\protect\ensuremath{\bigg(
-i\frac{\delta }{\delta A_{\mathrm{ext}}}
\Delta_{\text{R}}
\frac{\delta }{\delta \zeta }
 \bigg)}p^{\mathrm{I}}\protect\ensuremath{\big(
 J\big | A_{\mathrm{ext}}
 \big)} 
\exp\protect\ensuremath{\big(i \zeta J
 \big)} 
\\ 
= \exp\protect\ensuremath{\big(i \zeta J
 \big)}\exp\protect\ensuremath{\bigg(
\frac{\delta }{\delta A_{\mathrm{ext}}}
\Delta_{\text{R}}
 J
 \bigg)}p^{\mathrm{I}}\protect\ensuremath{\big(
 J\big | A_{\mathrm{ext}}
 \big)}
\\ 
= \exp\protect\ensuremath{\big(i \zeta J
 \big)}p^{\mathrm{I}}\protect\ensuremath{\big(
 J\big | A_{\mathrm{ext}} + \Delta_{\text{R}} J 
 \big)} . 
\hspace{0.4\columnwidth}\hspace{-0.4\twocolumnwidth} 
\label{eq:85ZF} % \nonumber % \Z 
\end{multline}}%
%+++++++++++++++++++++++++++++++++++++++++++++
We use condensed notation (\ref{eq:78NW}). 
The first step in (\ref{eq:85ZF}) is trivial; the second one is an application of a functional shift operator. This way, 
%=============================================
{\begin{multline}\hspace{0.4\columnwidth}\hspace{-0.4\twocolumnwidth} 
\Phi_{\mathrm{dev}}\protect\ensuremath{\big(
\zeta \big| A_{\mathrm{ext}}
 \big)} = 
\prod_{x,t}\bigg\{\int d J(x,t)\bigg\} 
\\ \times 
p^{\mathrm{I}}\protect\ensuremath{\big(
 J\big | A_{\mathrm{ext}} + \Delta_{\text{R}} J
 \big)} 
\exp\protect\ensuremath{\big(i \zeta J \big)} 
, 
\hspace{0.4\columnwidth}\hspace{-0.4\twocolumnwidth} 
\label{eq:86ZH} % \nonumber % \Z 
\end{multline}}%
%+++++++++++++++++++++++++++++++++++++++++++++
and 
%=============================================
\protect{\begin{align}{{
 \begin{aligned} 
p\protect\ensuremath{\big(
 J\big | A_{\mathrm{ext}}
 \big)} 
= p^{\mathrm{I}}\protect\ensuremath{\big(
 J\big | A_{\mathrm{ext}} + \Delta_{\text{R}} J
 \big)} . 
\end{aligned}}}%
\label{eq:87ZJ} % \nonumber % \Z 
\end{align}}%
%+++++++++++++++++++++++++++++++++++++++++++++
The classical content of this relation is crystal clear. Functional $p\protect\ensuremath{\big(
 J\big | A_{\mathrm{ext}}
 \big)}$ describes statistical properties of the current $ J(x,t)$ conditional on the external (macroscopic) field $ A_{\mathrm{ext}}(x,t)$. Functional $p^{\mathrm{I}}\protect\ensuremath{\big(
 J\big | A_{\mathrm{loc}}
 \big)}$ describes statistical properties of the current $ J(x,t)$ conditional on the local (microscopic) field $ A_{\mathrm{loc}}(x,t)$. The latter equals $ A_{\mathrm{ext}}(t)$ plus self-radiation of the current, 
%=============================================
{\begin{multline}\hspace{0.4\columnwidth}\hspace{-0.4\twocolumnwidth} 
 A_{\mathrm{loc}}(x,t) = A_{\mathrm{ext}}(x,t)
\\ 
+\int dx'dt' \Delta_{\text{R}}(x,x',t-t') J(x',t') . 
\hspace{0.4\columnwidth}\hspace{-0.4\twocolumnwidth} 
\label{eq:88ZK} % \nonumber % \Z 
\end{multline}}%
%+++++++++++++++++++++++++++++++++++++++++++++
In quantum electrodynamics, this interpretation applies with replacement of ``statistical'' by ``quasistatistical.'' 

In the general case, the dressing relation for P-functionals reads, 
%\begin{widetext} 
%=============================================
{\begin{multline}\hspace{0.4\columnwidth}\hspace{-0.4\twocolumnwidth} 
p\protect\ensuremath{\big(
 J,D,D^*\big | A_{\mathrm{ext}},E_{\mathrm{ext}},E_{\mathrm{ext}}^*
 \big)} \\ 
= p^{\mathrm{I}}\protect\ensuremath{\big(
 J,D,D^*\big | A_{\mathrm{ext}} + \Delta_{\text{R}} J, 
 E_{\mathrm{ext}}+G_{\text{R}} D,E_{\mathrm{ext}}^*+G_{\text{R}}^* D^*
 \big)} , 
\hspace{0.4\columnwidth}\hspace{-0.4\twocolumnwidth} 
\label{eq:57HR} % \nonumber % \Z 
\end{multline}}%
%+++++++++++++++++++++++++++++++++++++++++++++
where the functional $p_{\text{I}}\protect\ensuremath{\big(
 J,D,D^*\big | A_{\mathrm{ext}},E_{\mathrm{ext}},E_{\mathrm{ext}}^*
 \big)}$ is given by the relation, 
%=============================================
{\begin{multline}\hspace{0.4\columnwidth}\hspace{-0.4\twocolumnwidth} 
p^{\mathrm{I}}\protect\ensuremath{\big(
 J,D,D^*\big | A_{\mathrm{ext}},E_{\mathrm{ext}},E_{\mathrm{ext}}^*
 \big)} \\ 
= \prod_{x,t}\bigg\{\frac{(dxdt)^3}{2\pi ^3}\int d\zeta (x,t)d^2\nu (x,t)\bigg\}
\text{Tr} \hat \rho _{\textrm{dev}} 
\\ \times 
{\mathcal T}{\mbox{\rm\boldmath$:$}}\exp \protect\protect\ensuremath{\big[
i \zeta\protect\ensuremath{\big(
\hat J' - J
 \big)} 
-i\mu^*\protect\ensuremath{\big(
\hat D' - D
 \big)} +i\mu\protect\ensuremath{\big(
\hat D^{\prime\dag} - D^*
 \big)} 
\big]} 
{\mbox{\rm\boldmath$:$}}
. 
\hspace{0.4\columnwidth}\hspace{-0.4\twocolumnwidth} 
\label{eq:58HS} % \nonumber % \Z 
\end{multline}}%
%+++++++++++++++++++++++++++++++++++++++++++++
%\end{widetext}%
The primed operators were defined in section \ref{ch:UPS}. 
Derivation and interpretation of eq.\ (\protect\ref{eq:57HR}) are no different from those of eq.\ (\protect\ref{eq:87ZJ}). 
%********************************************************
\section{Discussion: causality and regularisations}%
\label{ch:R}
%********************************************************
%\subsection{Causality as a consistency condition}%
%\chlabel{ch:RC}
%********************************************************
It should not be overlooked that eq.\ (\protect\ref{eq:87ZJ}) is consistent only due to causality properties of P-functionals. Here is a simple example. Assume that all quantities in (\ref{eq:87ZJ}) do not depend on time. The external field shifts the Gaussian distribution of the current, 
%=============================================
\protect{\begin{align}{{
 \begin{aligned} 
p^{\mathrm{I}}\protect\ensuremath{\big(
 J \big | A_{\mathrm{e}} 
 \big)} 
= \frac{1}{\sqrt{2\pi}\, J_0 }
\exp\protect\protect\ensuremath{\bigg[
-\frac{\protect\ensuremath{\big(
 J - \chi A_{\mathrm{e}} \big)}^2 
}{2 J_0^2}
\bigg]} , 
\end{aligned}}}%
\label{eq:3ZZ} % \nonumber % \Z 
\end{align}}%
%+++++++++++++++++++++++++++++++++++++++++++++
where $ J_0>0$ and $\chi$ are real constants. In place of (\ref{eq:88ZK}) we postulate a scalar formula, 
%=============================================
\protect{\begin{align}{{
 \begin{aligned} 
 A_{\mathrm{loc}} = A_{\mathrm{e}} + \Delta_{\text{R}} J, 
\end{aligned}}}%
\label{eq:4AA} % \nonumber % \Z 
\end{align}}%
%+++++++++++++++++++++++++++++++++++++++++++++
where $\Delta_{\text{R}}$ is one more real constant. For the ``dressed current'' we find, 
%=============================================
\protect{\begin{align}{{
 \begin{aligned} 
p\protect\ensuremath{\big(
 J \big | A_{\mathrm{e}} 
 \big)} 
= \frac{1}{\sqrt{2\pi}\, J_0 }
\exp\protect\protect\ensuremath{\bigg[
-\frac{\protect\ensuremath{\big(
 J - \chi \Delta_{\text{R}} J - \chi A_{\mathrm{e}} \big)}^2 
}{2 J_0^2}
\bigg]} . 
\end{aligned}}}%
\label{eq:5AB} % \nonumber % \Z 
\end{align}}%
%+++++++++++++++++++++++++++++++++++++++++++++
This function is not normalised, 
%=============================================
\protect{\begin{align}{{
 \begin{aligned} 
\int d J \, p\protect\ensuremath{\big(
 J \big | A_{\mathrm{e}} 
 \big)} = \frac{1}{1-\chi \Delta_{\text{R}}} \neq 1, 
\end{aligned}}}%
\label{eq:6AC} % \nonumber % \Z 
\end{align}}%
%+++++++++++++++++++++++++++++++++++++++++++++
and cannot be a probability distribution for anything. 

To see how causality breaks this vicious circle of same-time interactions consider another simple example. We generalise (\ref{eq:3ZZ}) to two currents, 
%=============================================
{\begin{multline}\hspace{0.4\columnwidth}\hspace{-0.4\twocolumnwidth} 
p^{\mathrm{I}}\protect\ensuremath{\big(
 J , J' \big | A_{\mathrm{e}}, A_{\mathrm{e}}' 
 \big)} 
\\ 
= \frac{1}{2\pi J_0^2 }
\exp\protect\protect\ensuremath{\bigg[
-\frac{
\protect\ensuremath{\big( J - \chi A_{\mathrm{e}} \big)}^2
+ 
\protect\ensuremath{\big( J' - \chi A_{\mathrm{e}}' \big)}^2 
}{2 J_0^2}
\bigg]} . 
\hspace{0.4\columnwidth}\hspace{-0.4\twocolumnwidth} 
\label{eq:7AD} % \nonumber % \Z 
\end{multline}}%
%+++++++++++++++++++++++++++++++++++++++++++++
The primed current preceeds the unprimed one in time; therefore it may affect the latter but not {\em vice versa\/}. Same-time interactions are not allowed either. The simplest case of such interaction is, 
%=============================================
\protect{\begin{align}{{
 \begin{aligned} 
& A_{\mathrm{loc}} = A_{\mathrm{e}} + \Delta_{\text{R}} J', & 
& A_{\mathrm{loc}}' = A_{\mathrm{e}}' . 
\end{aligned}}}%
\label{eq:10AH} % \nonumber % \Z 
\end{align}}%
%+++++++++++++++++++++++++++++++++++++++++++++
For the dressed currents we then have, 
%=============================================
{\begin{multline}\hspace{0.4\columnwidth}\hspace{-0.4\twocolumnwidth} 
p\protect\ensuremath{\big(
 J , J' \big | A_{\mathrm{e}}, A_{\mathrm{e}}' 
 \big)} 
= \frac{1}{2\pi J_0^2 }
\\ \times
\exp\protect\protect\ensuremath{\bigg[
-\frac{
\protect\ensuremath{\big( J - \chi \Delta_{\text{R}} J' - \chi A_{\mathrm{e}} \big)}^2
+ 
\protect\ensuremath{\big( J' - \chi A_{\mathrm{e}}' \big)}^2 
}{2 J_0^2}
\bigg]} . 
\hspace{0.4\columnwidth}\hspace{-0.4\twocolumnwidth} 
\label{eq:8AE} % \nonumber % \Z 
\end{multline}}%
%+++++++++++++++++++++++++++++++++++++++++++++
Unlike (\ref{eq:5AB}), this function is both positive and normalised, 
%=============================================
\protect{\begin{align}{{
 \begin{aligned} 
\int d J'
\int d J\,p\protect\ensuremath{\big(
 J , J' \big | A_{\mathrm{e}}, A_{\mathrm{e}}' 
 \big)} = 1 . 
\end{aligned}}}%
\label{eq:9AF} % \nonumber % \Z 
\end{align}}%
%+++++++++++++++++++++++++++++++++++++++++++++
It is therefore a genuine two-dimensional conditional probability distribution for a correlated pair of currents. 

The order of integrations in (\ref{eq:9AF}) is chosen so as to make the result obvious. Indeed, (\ref{eq:8AE}) has the structure, 
%=============================================
\protect{\begin{align}{{
 \begin{aligned} 
p\protect\ensuremath{\big(
 J , J' \big | A_{\mathrm{e}}, A_{\mathrm{e}}' 
 \big)} = p\protect\ensuremath{\big(
 J\big | A_{\mathrm{e}} , J' 
 \big)}p'\protect\ensuremath{\big(
 J' \big | A_{\mathrm{e}}' 
 \big)} , 
\end{aligned}}}%
\label{eq:12AK} % \nonumber % \Z 
\end{align}}%
%+++++++++++++++++++++++++++++++++++++++++++++
where 
%=============================================
\protect{\begin{align}{{
 \begin{aligned} 
p\protect\ensuremath{\big(
 J\big | A_{\mathrm{e}} , J' 
 \big)} & = \frac{1}{\sqrt{2\pi}\, J_0 }
%\\ \times
\exp\protect\protect\ensuremath{\bigg[
-\frac{
\protect\ensuremath{\big( J - \chi \Delta_{\text{R}} J' - \chi A_{\mathrm{e}} \big)}^2
}{2 J_0^2}
\bigg]} , \\ 
p'\protect\ensuremath{\big(
 J' \big | A_{\mathrm{e}}' 
 \big)} & = \frac{1}{\sqrt{2\pi}\, J_0 }
%\\ \times
\exp\protect\protect\ensuremath{\bigg[
-\frac{
\protect\ensuremath{\big( J' - \chi A_{\mathrm{e}}' \big)}^2 
}{2 J_0^2}
\bigg]} . 
\end{aligned}}}%
\label{eq:13AL} % \nonumber % \Z 
\end{align}}%
%+++++++++++++++++++++++++++++++++++++++++++++
The later current is conditional on the earlier one and the external field. The earlier current is conditional only on the external field. 
% hide 
Similar structures should emerge for any time sequence of currents irrespective of any detals of the interaction. The only requirement is that each current depends only on those preceding it in time. 

In real problems with continuous time, critical for cancellation of same-time interactions are {\em regularisations\/}. So, in \mbox{Ref.\ \protect\cite{BWO}}, {\em causal regularisation\/} was applied to the retarded Green function of the emerging equation for phase-space amplitudes. This made noise sources present in the said equation independent of the amplitudes at the same time, resulting in the Ito calculus being chosen. The effect of causal regularisation is thus twofold: to introduce an infinitesimal delay into the phase-space equation, which is in essence time discretisation, and to prevent same-time interactions. A conceptual connection between the above simple examples and the causal regularisation is obvious. For a discussion of the connection between causal regularisation and suppression of infinities in relativistic quantum field theory we refer the reader to appendix D 
of \mbox{Ref.\ \protect\cite{WickCaus}}. 

% hide
%********************************************************
\section{Conclusion and outlook}%
%********************************************************
It is shown that phase-space concepts such as time-normal operator ordering and P-functonal provide a natural framework for quantum interactions of light and matter. In a forthcoming paper \cite{Sudar} this framework will be extended to macroscopic interactions of distinguishable devices. 
%********************************************************
\section{Acknowledgements}%
%********************************************************
Support of SFB/TRR 21 and of the Humboldt Foundation is gratefully acknowledged. 

%********************************************************
\appendix*
\section{Functional probability distributions and inversion formulae}%
\label{ch:TNCF}
%********************************************************
The goal of this appendix is to derive the inversion formula (\ref{eq:98EC}), and to point to mathematical complications hidden behind apparent simplicity of our formulae. For simplicity we consider a c-number random quantity $J(t)$. For general non-Markovian systems, classical statistical averaging is formally a functional (path) integral, 
%=============================================
\protect{\begin{align}{{
 \begin{aligned} 
\overline{\hspace{0.1ex} J(t_1)\cdots J(t_m)\hspace{0.1ex}} 
= \prod_{t}\bigg\{\int d J(t)\bigg\}\,p\protect\ensuremath{(
 J
 )}\, J(t_1)\cdots J(t_m) , 
\end{aligned}}}%
\label{eq:51DP} % \nonumber % \Z 
\end{align}}%
%+++++++++++++++++++++++++++++++++++++++++++++
where $p\protect\ensuremath{(
 J
 )}$ is a functional probability distribution over the random functions (paths) $ J(t)$. We emphasise that $p\protect\ensuremath{(
 J
 )}$ is not a function of variable $ J$, but a functional of a function $ J(t)$. 

For all practical purposes, $t$ in (\ref{eq:51DP}) may be thought of as a discrete index. Functional integration is then regarded a multiple integration over variables $ J(t)= J_t$, each defined in an infinitely narrow Trotter time slice $
\Delta t$ labelled by index $t$. With this simplified view, algebraic manipulation of the path integral becomes straightforward. For example, let us express $p\protect\ensuremath{(
 J
 )}$ in terms of the characteritic functional of quantum averages (\ref{eq:51DP}),
%=============================================
{\begin{multline}\hspace{0.4\columnwidth}\hspace{-0.4\twocolumnwidth} 
\Phi\protect\ensuremath{\big(
\zeta 
 \big)} = \overline{\hspace{0.1ex}\exp \protect\protect\ensuremath{\bigg[
i \int dt \zeta (t) J(t)
\bigg]} \hspace{0.1ex}} \\ 
= \prod_{t}\bigg\{\int d J(t)\bigg\}p( J)\exp \protect\protect\ensuremath{\bigg[
i \int dt \zeta (t) J(t)
\bigg]} . 
\hspace{0.4\columnwidth}\hspace{-0.4\twocolumnwidth} 
\label{eq:52DQ} % \nonumber % \Z 
\end{multline}}%
%+++++++++++++++++++++++++++++++++++++++++++++
Thinking discretised time we replace, 
%=============================================
\protect{\begin{align}{{
 \begin{aligned} 
\exp \protect\protect\ensuremath{\bigg[
i \int dt \zeta (t) J(t)
\bigg]} \to \prod_t \mathrm{e}^{i\Delta t \zeta_t J_t} . 
\end{aligned}}}%
\label{eq:53DR} % \nonumber % \Z 
\end{align}}%
%+++++++++++++++++++++++++++++++++++++++++++++
The discretised approximation to (\ref{eq:52DQ}) reads, 
%=============================================
\protect{\begin{align}{{
 \begin{aligned} 
\Phi\protect\ensuremath{\big(
\zeta 
 \big)} = 
\prod_{t}\bigg\{\int \mathrm{e}^{i\Delta t \zeta_t J_t}d J_t\bigg\}\,p\protect\ensuremath{(
 J
 )} . 
\end{aligned}}}%
\label{eq:54DS} % \nonumber % \Z 
\end{align}}%
%+++++++++++++++++++++++++++++++++++++++++++++
This is nothing but a multidimensional Fourier-transform. Inverting it ``slicewise'' we find, 
%=============================================
{\begin{multline}\hspace{0.4\columnwidth}\hspace{-0.4\twocolumnwidth} 
p\protect\ensuremath{(
 J
 )} = \prod_{t}\bigg\{\frac{\Delta t}{2\pi }\int \mathrm{e}^{-i\Delta t \zeta_t J_t}d\zeta_t\bigg\}
\Phi\protect\ensuremath{\big(
\zeta 
 \big)} 
\\ 
= \prod_{t}\bigg\{\frac{\Delta t}{2\pi }\int d\zeta_t \bigg\}
\Phi\protect\ensuremath{\big(
\zeta 
 \big)} \exp\protect\protect\ensuremath{\Big[
-i\Delta t \sum_t\zeta_t J_t
\Big]} 
.
\hspace{0.4\columnwidth}\hspace{-0.4\twocolumnwidth} 
\label{eq:59DX} % \nonumber % \Z 
\end{multline}}%
%+++++++++++++++++++++++++++++++++++++++++++++
Restoring continuity of time, we obtain the inversion formula, 
%=============================================
\protect{\begin{align}{{
 \begin{aligned} 
p\protect\ensuremath{(
 J
 )} = \prod_{t}\bigg\{\frac{dt}{2\pi }\int d \zeta (t)\bigg\} \exp \protect\protect\ensuremath{\bigg[
- i \int dt \zeta (t) J(t)
\bigg]}\Phi\protect\ensuremath{\big(
\zeta 
 \big)} .
\end{aligned}}}%
\label{eq:55DT} % \nonumber % \Z 
\end{align}}%
%+++++++++++++++++++++++++++++++++++++++++++++
The infinitesimal factor $\prod_t dt$ leaves no doubt that this expression is only symbolic. 

It should be noted that the notation we use for path integrals is equally symbolic. As an example, consider a well defined mathematical concept: the Wiener process. In discretised time, the probability density for the Wiener process reads, 
%=============================================
\protect{\begin{align}{{
 \begin{aligned} 
\prod_t \protect\protect\ensuremath{\bigg\{
\frac{d\Delta J_t}{\sqrt{2\pi\Delta t}}
\exp\protect\protect\ensuremath{\bigg[
-\frac{\protect\ensuremath{\big(
\Delta J_t
 \big)}^2 }{2\Delta t}
\bigg]}
\bigg\}} , 
\end{aligned}}}%
\label{eq:71JF} % \nonumber % \Z 
\end{align}}%
%+++++++++++++++++++++++++++++++++++++++++++++
where $\Delta J_t = J_t-J_{t-\Delta t}$ are stochastic increments. 
From the first glance, continuous limit may be achieved introducing the discretised derivative, 
%=============================================
\protect{\begin{align}{{
 \begin{aligned} 
J'_t = \frac{\Delta J_t}{\Delta t}. 
\end{aligned}}}%
\label{eq:69JD} % \nonumber % \Z 
\end{align}}%
%+++++++++++++++++++++++++++++++++++++++++++++
so that
%=============================================
{\begin{multline}\hspace{0.4\columnwidth}\hspace{-0.4\twocolumnwidth} 
\prod_t \protect\protect\ensuremath{\bigg\{
\frac{d\Delta J_t}{\sqrt{2\pi\Delta t}}
\exp\protect\protect\ensuremath{\bigg[
-\frac{\protect\ensuremath{\big(
\Delta J_t
 \big)}^2 }{2\Delta t}
\bigg]}
\bigg\}} \\ 
= \prod_t \protect\protect\ensuremath{\bigg[
d J'_t\sqrt{\frac{\Delta t}{2\pi}}
\bigg]} 
\exp\protect\protect\ensuremath{\bigg[
-\sum_t\frac{\Delta t\protect\ensuremath{\big(
 J'_t
 \big)}^2 }{2}
\bigg]}
. 
\hspace{0.4\columnwidth}\hspace{-0.4\twocolumnwidth} 
\label{eq:68JC} % \nonumber % \Z 
\end{multline}}%
%+++++++++++++++++++++++++++++++++++++++++++++
In the continuous limit, we have for the probability density, 
%=============================================
\protect{\begin{align}{{
 \begin{aligned} 
\prod_t \protect\protect\ensuremath{\bigg[
d J'(t)\sqrt{\frac{d t}{2\pi}}
\bigg]} 
\exp\protect\protect\ensuremath{\bigg\{
-\frac{1}{2}\int dt\protect\protect\ensuremath{\big[
 J'(t)
\big]}^2
\bigg\}}
. 
\end{aligned}}}%
\label{eq:70JE} % \nonumber % \Z 
\end{align}}%
%+++++++++++++++++++++++++++++++++++++++++++++
Infinitesimal scaling factors are indeed eliminated from the exponent, but the overall factor $\prod_t \sqrt{dt}$ persists. Because of this factor, quantity (\ref{eq:70JE}) is zero for all functions for which the integral in the exponent is defined (as expected). For nondifferentiable functions --- which are of actual interest --- eq.\ (\protect\ref{eq:70JE}) is useless, because anyway it has to be specified through some limiting procedure. Hence (\ref{eq:70JE}) is no more than a symbolic way of writing the discretised approximation (\ref{eq:71JF}).

%******************************************* 

\begin{thebibliography}{27}
\bibitem[{end()}]{endFerm}
\bibinfo{note}{The intermediate and the final expressions in Eq.\
  (\ref{eq:47DK}) differ in an even permutation of atomic operators, so that
  this formula holds irrespective of whether $\hat\psi_{g},\hat\psi_{e}$ are
  bosonic or fermionic. For a discussion of orderings of fermionic operators
  see \cite{APIII}}.
\end{thebibliography}
\end{document}